\documentclass[prc,aps,twocolumn,showpacs,amssymb,superscriptaddress,fleqn]{revtex4-1}

\usepackage{amsmath}
\usepackage{color}
\usepackage{graphicx}
\usepackage{dcolumn}
\usepackage{mathtools}
\usepackage{relsize}
\usepackage{float}
\usepackage{braket}

\newcommand{\lwk}{{{\rm low}\mbox{-}k}}
\newcommand{\vlwk}{$V_{{\rm low}\mbox{-}k}$}

\newcommand{\zbb}{$0\nu\beta\beta$}
\newcommand{\dbb}{$2\nu\beta\beta$}
\newcommand{\heff}{$H_{\rm eff}$}
\newcommand{\heffs}{$H_{\rm eff}$s}
\newcommand{\qbox}{$\hat{Q}$~box}
\newcommand{\nme}{$M^{0\nu}$}
\newcommand{\nmes}{$M^{0\nu}$s}
\newcommand{\nmed}{$M^{2\nu}$}
\newcommand{\nmeds}{$M^{2\nu}$s}

\begin{document}

\title{The calculation of the neutrinoless double-$\beta$ decay matrix
  element within the realistic shell model}

\author{L. Coraggio}
\affiliation{Istituto Nazionale di Fisica Nucleare, \\
Complesso Universitario di Monte  S. Angelo, Via Cintia - I-80126 Napoli, Italy}
\author{A. Gargano}
\affiliation{Istituto Nazionale di Fisica Nucleare, \\
Complesso Universitario di Monte  S. Angelo, Via Cintia - I-80126 Napoli, Italy}
\author{N. Itaco}
\affiliation{Dipartimento di Matematica e Fisica, Universit\`a degli
  Studi della Campania ``Luigi Vanvitelli'', viale Abramo Lincoln 5 -
  I-81100 Caserta, Italy}
\affiliation{Istituto Nazionale di Fisica Nucleare, \\ 
Complesso Universitario di Monte  S. Angelo, Via Cintia - I-80126 Napoli, Italy}
\author{R. Mancino}
\affiliation{Dipartimento di Matematica e Fisica, Universit\`a degli
  Studi della Campania ``Luigi Vanvitelli'', viale Abramo Lincoln 5 -
  I-81100 Caserta, Italy}
\affiliation{Istituto Nazionale di Fisica Nucleare, \\ 
Complesso Universitario di Monte  S. Angelo, Via Cintia - I-80126 Napoli, Italy}
\author{F. Nowacki}
\affiliation{Universit\'e de Strasbourg, IPHC, 23 rue du Loess 67037 Strasbourg, France}
\affiliation{CNRS, IPHC UMR 7178, 67037 Strasbourg, France}
\affiliation{Dipartimento di Matematica e Fisica, Universit\`a degli
  Studi della Campania ``Luigi Vanvitelli'', viale Abramo Lincoln 5 -
  I-81100 Caserta, Italy}

\begin{abstract}
We approach the calculation of the nuclear matrix element of the
neutrinoless double-$\beta$ decay process, considering the
light-neutrino-exchange channel, by way of the realistic shell model.
To this end, we start from a realistic nucleon-nucleon potential and
then derive the effective shell-model Hamiltonian and \zbb -decay
operator within the many-body perturbation theory.
We focus on investigating the perturbative properties of the effective
shell-model operator of such a decay process, aiming to establish the
degree of reliability of our predictions.
The contributions of the so-called short-range correlations and of the
correction of Pauli-principle violations to the effective shell-model
operator, the latter introduced in many-valence nucleon systems, are
also taken into account.
The subjects of our study are a few candidates to the \zbb -decay
detection, in a mass interval ranging from $A=48$ up to $A=136$, whose
spin- and spin-isospin-dependent decay properties we have studied in
previous works.
Our results evidence that the effect of the
  renormalization of the \zbb-beta decay operator on the values of the
  nuclear matrix elements is less relevant than what we have obtained
  in previous studies of the effective single-body GT transitions
  operating also in the two neutrinos double-beta decay
\end{abstract}

\pacs{21.60.Cs, 21.30.Fe, 27.60.+j, 23.40-s}

\maketitle

\section{Introduction}
\label{intro}
The search for evidence of the neutrinoless double-$\beta$ decay
(\zbb) is at present one of the major goal in experimental
physics, because of its implications in our understanding of both the
limits of Standard Model and the intrinsic nature of the neutrino
\cite{Avignone08,DellOro15,Henning16}.

Double-$\beta$ decay is a second-order electroweak process and the
rarest among the nuclear transitions.
As is well known, the $\beta$-decay with the emission of two
neutrinos  (\dbb) occurs with half-lives that exceed
$10^{18}$~yr, and, at present, the strongest limits on \zbb~decay have
been set as $> 0.9 \times 10^{26}$~yr for $^{76}$Ge decay by GERDA
experiment \cite{GERDA19}, $> 4 \times 10^{24}$~yr for $^{130}$Te
decay by CUORE experiment \cite{CUORE}, and $> 1.07 \times 10^{26}$~yr
for $^{136}$Xe decay by KamLAND-ZEN collaboration \cite{KamLAND16}.

The detection of \zbb~decay - a process that requires the violation of
the conservation of the lepton number - would pave the way to a
physics whose mechanisms should lie beyond the Standard Model.
It would reveal the nature of the massive neutrino as a Majorana
rather than a Dirac particle, being a spin-$\frac{1}{2}$ particle
which coincides with its anti-particle.
Such a feature may provide insight into the matter-antimatter
asymmetry in the Universe within the framework of CP violation effects
due to the see-saw and leptogenesis mechanisms \cite{Buchmuller05}.

  The half-life of the \zbb~decay may be expressed as
  
\begin{equation}
\left[ T^{0\nu}_{1/2}\right]^{-1} = G^{0\nu} \left| M^{0\nu} \right|^2
\left| f(m_i,U_{ei})\right|^2~,
\label{halflife}
\end{equation}

where $G^{0\nu}$ is the so-called phase-space factor (or kinematic
factor) \cite{Kotila12,Kotila13}, \nme~is the nuclear matrix element (NME) directly
related to the wave functions of the parent and grand-daughter nuclei,
and $f(m_i,U_{ei})$ accounts for the physics beyond the standard model
through the neutrino masses $m_i$ and their mixing matrix elements
$U_{ei}$.
The explicit form of $f(m_i,U_{ei})$ depends on the adopted model of
\zbb~decay (light- and/or heavy neutrino exchange, Majoron exchange,
...).

In particular, within the mechanisms of light-neutrino
  exchange, $f(m_i,U_{ei})$ has the following expression:
\[
f(m_i,U_{ei}) = g_A^2 \frac{ \langle m_{\nu}\rangle}{m_e}
\]
\noindent
where $g_A$ is the axial coupling constant, $m_e$ is the electron
mass, and $\langle m _{\nu} \rangle = \sum_i (U_{ei})^2 m_i$ is the
effective neutrino mass.
The above expression and the one in (\ref{halflife}) evidence the
pivotal role played by the calculation of the NME.

In fact, a reliable estimate of its value provides important
information about some crucial issues, since the neutrino effective
mass can be expressed in terms of $M^{0 \nu}$, of the half-life
$T^{0\nu}_{1/2}$, and the so-called nuclear structure factor
$F_N=G^{0\nu} \left| M^{0\nu} \right|^2g _{A}^4$, as $\langle m _{\nu}
\rangle = m_e\left[ F_N T^{0\nu}_{1/2}\right]^{-1/2}$.
Moreover, combining the calculated nuclear structure factor with
neutrino mixing parameters \cite{PDG18} and limits on $\langle m
_{\nu} \rangle$ from current experiments, one may extract an
estimation of the half-life an experiment should measure in order to
be sensitive to a particular value of the neutrino effective mass
\cite{Avignone08}.
All the above considerations evidence that reliable calculations of $M^{0
  \nu}$ are of paramount importance, and, currently, various nuclear
structure models are employed to study this process.

It should be noted that currently {\it ab initio} calculations may be
carried out only for light nuclei \cite{Pastore18}, but the parameters
that locate the best candidates of experimental interest - $Q$-value and
phase-space factor of the decay, and the isotopic abundance of the
parent nucleus - point to the region of medium- and heavy-mass
nuclei.
This is why the nuclear structure models which are mostly employed to
study the \zbb~decay of nuclei of experimental interest
are the Interacting Boson Model (IBM) \cite{Barea09,Barea12,Barea13},
the Quasiparticle Random-Phase Approximation (QRPA)
\cite{Simkovic09,Fang11,Faessler12}, Energy Density
Functional methods \cite{Rodriguez10}, the Covariant
Density Functional Theory \cite{Song14,Yao15,Song17}, the
Generator-Coordinate Method (GCM) \cite{Jiao17,Yao18,Jiao18,Jiao19},
and the Shell Model (SM)
\cite{Menendez09a,Menendez09b,Horoi13b,Neacsu15,Brown15}.

Actually, for all these calculations the truncation of the full
Hilbert space is mandatory to diagonalize the nuclear Hamiltonian, and
the renormalization of the effective Hamiltonian \heff~ is
generally performed fitting the parameters which characterize each
model, and that are associated with the degrees of freedom of the reduced
model space, to some spectroscopic properties of the nuclei under
investigation.
As a matter of fact, the determination of the model parameters is
carried out taking into account the specific abilities of the
different models, leading to a spread of the results of the
calculation of \nme~with different approaches, which agree
within a factor $\sim 2 \div 3$ (see, for example, Fig. 5 in
Ref. \cite{Engel17} and references therein).

Aside from the renormalization of the nuclear Hamiltonian, whenever
transition properties are calculated using wave functions obtained
diagonalizing \heff, the free constants that appear in the definition
of the decay operators - proton and neutron electric charges, spin and
orbital gyromagnetic factors, etc. - need also to be modified to
account for the degrees of freedom that do not appear explicitly
because of the truncation of the full Hilbert space, a procedure that
is also performed by fitting them to reproduce observables.

This leads to the well-known problem of quenching the free value of the
axial coupling constant $g_A^{free}=1.2723$ \cite{PDG18} via a
quenching factor $q$ \cite{Towner87}, whose choice depends on the
nuclear structure model, the dimensions of the reduced Hilbert space,
and the mass of the nuclei under investigation \cite{Suhonen17b}.
We point out that both the renormalization of many-body correlations
and the corrections due to the subnucleonic structure of the nucleons
\cite{Park93,Pastore09,Piarulli13,Baroni16b} are needed to handle the
issue of quenching $g_A$, whose $q$ value is commonly determined
fitting experimental data from Gamow-Teller (GT) transitions.

In this work, we will tackle the problem to calculate the
\zbb -decay NME, in the light-neutrino-exchange channel, within the
framework of the realistic shell model (RSM) \cite{Coraggio09a}, that
is the effective shell-model Hamiltonian \heff~ and decay operators are
consistently derived starting from a realistic nucleon-nucleon ($NN$)
potential $V_{NN}$.
This can be achieved by way of many-body perturbation theory
\cite{Kuo95,Hjorth95,Suzuki95,Coraggio12a}, in order to construct
single-particle (SP) energies, two-body matrix elements of the
residual interaction (TBME), and two-body matrix elements of the
effective \zbb -decay operator ($O^{0\nu}_{\rm eff}$) within
a microscopic approach.

Our input $V_{NN}$ is the high-precision CD-Bonn $NN$ potential
\cite{Machleidt01b}, whose repulsive high-momentum components are
smoothed out using the \vlwk~approach \cite{Bogner02}.
The \vlwk~procedure provides a consistent treatment of the so-called
short-range correlations, which need to be included since the
basis that is employed for the perturbative expansion consists of
uncorrelated wave functions that do not vanish in the short range,
the latter corresponding to the region of the strong-repulsive
components of $V_{NN}$ \cite{Bethe71,Kortelainen07}.

We have pursued the approach of RSM already to calculate the
\dbb~ NME (\nmed) in
Refs. \cite{Coraggio17a,Coraggio19a} for $^{48}$Ca, $^{76}$Ge,
$^{82}$Se, $^{130}$Te, and $^{136}$Xe.
In the same works a study of the spectroscopy of these nuclei,
focussing in particular on spin-dependent decays, and of their
GT-strength distributions has been also reported.

It should be mentioned that similar SM calculations of \nme~have
been performed in Refs. \cite{Wu85,Song91,Holt13d}.

Calculations of \nme~for $^{48}$Ca decay in
Refs. \cite{Wu85,Song91} are based on a consistent treatement of the
short-range correlations in terms of the defect wave functions
\cite{Bethe71} derived from the calculation of the reaction matrix $G$
from Paris \cite{Lacombe80} and Reid \cite{Reid68} $NN$ potentials.
The $G$-matrix vertices appear also in the perturbative expansion of
\heff~ and $O^{0\nu}_{\rm eff}$, which is arrested at second order.

The authors in Ref. \cite{Holt13d} focus their attention on \zbb~decay
of $^{76}$Ge and $^{82}$Se, and start from a chiral N$^{3}$LO $NN$
potential \cite{Entem02} renormalized through the \vlwk~technique.
They expand $O^{0\nu}_{\rm eff}$ up to third order in perturbation
theory, and include the effects of high-momentum (short-range)
correlations via an effective Jastrow function that has been fit to
the results of Brueckner-theory calculations \cite{Simkovic09}.
The wave functions of parent and grand-daughter nuclei have been
calculated using two different phenomenological SM \heffs~ from
Refs. \cite{Horoi13c,Menendez09b} for $^{76}$Ge and $^{82}$Se decays,
respectively.

In Section \ref{outline} details about the derivation of the
effective SM Hamiltonian and \zbb~operator from a realistic $V_{NN}$
are reported.
We show the results of calculations of \nme~for  $^{48}$Ca, $^{76}$Ge,
$^{82}$Se, $^{130}$Te, and $^{136}$Xe double-$\beta$ decay in Section
\ref{calculations}, and present also a detailed study of the
perturbative properties of $O^{0\nu}_{\rm eff}$, together with an
analysis of the angular momentum-parity matrix-element distributions
and a comparison with recent SM results.
The conclusions of this study are drawn in Section \ref{conclusions},
together with the perspectives of our current project.

\section{Theoretical framework}\label{outline}
\subsection{The effective SM Hamiltonian}
The first step of our calculations is to consider the high-precision
CD-Bonn $NN$ potential \cite{Machleidt01b}.

The repulsive high-momentum components of CD-Bonn potential are
responsible of its non-perturbative behavior, and we renormalize them
in terms of the \vlwk~approach \cite{Bogner02,Coraggio09a}.
This renormalization of $V_{NN}$ occurs through a unitary
transformation $\Omega$, which decouples the full momentum space of
the two-nucleon Hamiltonian $H^{NN}$ into two subspaces; the first one
is associated to relative-momentum configurations below a cutoff
$\Lambda$ and a projector operator $P$, the second one is defined in
terms of its complement $Q=\mathbf{1}-P$ \cite{Coraggio19b}.
Obviously, as a unitary transformation, $\Omega$ preserves the
physics of the original potential for the two-nucleon system, namely
the calculated values of all $NN$ observables are the same as those
reproduced by $V_{NN}$.
 
This procedure does not affect the kinetic term $T$ of
  $H^{NN}$, $T$ being a one-body operator that already satisfies the
  decoupling condition between $P$ and $Q$ subspaces, and provides a
smooth potential $V_\lwk \equiv P \Omega V_{NN} \Omega^{-1} P$ which
is defined as equal to zero for momenta $k>\Lambda$, and that can be
employed as interaction vertex in perturbative many-body calculations.
We have chosen the value of the cutoff $\Lambda$, as in our previous
works \cite{Coraggio17a, Coraggio19a}, to be equal to $2.6$ fm$^{-1}$,
since the larger the cutoff the smaller is the role of the missing
induced three-nucleon force (3NF) \cite{Coraggio15b}.
The Coulomb potential is explicitly taken into account in the
proton-proton channel.

The \vlwk~can be now employed as the two-body interaction component of
the full nuclear Hamiltonian for $A$ interacting nucleons:

\begin{equation}
 H =  \sum_{i=1}^{A} \frac{p_i^2}{2m} + \sum_{i<j=1}^{A} V_\lwk^{ij}
 = T + V_\lwk ~.\label{htotal}
\end{equation}

\noindent
The diagonalization of the above $A-$body Hamiltonian within an
infinite Hilbert space is unfeasible, and in the shell model the
computational problem is reduced to the finite number of degrees of
freedom characterizing the physics of a limited number of interacting
nucleons, which can access only to the configurations of a model space
spanned by a few accessible orbitals.
This can be achieved by breaking up the Hamiltonian $H$ in
Eq. (\ref{htotal}), through an auxiliary one-body potential $U$, as a
sum of a one-body term $H_0$, whose eigenvectors set up the
shell-model basis, and a residual interaction $H_1$:

\begin{eqnarray}
 H &= & T + V_\lwk = (T+U)+(V_\lwk-U)= \nonumber \\
~& = &H_{0}+H_{1}~.\label{smham}
\end{eqnarray}

The following step is to derive an effective shell-model Hamiltonian
\heff, that takes into account the core polarization due to the
interaction between the valence nucleons and those belonging  to the
closed core, as well as the interaction between configurations
belonging to the model space and those corresponding the shells above
it.

As already mentioned, $H_{\rm eff}$ is derived by way of the many-body
perturbation theory, an approach developed by Kuo and coworkers
through the 1970s \cite{Kuo90,Kuo95}.
This is commonly known as the $\hat{Q}$-box-plus-folded-diagram method
\cite{Kuo71}, the $\hat{Q}$ box being a function of the unperturbed
energy $\epsilon$ of the valence particles:
\begin{equation}
\hat{Q}(\epsilon) = P H_1 P + P H_1 Q \frac{1}{\epsilon - QHQ} Q H_1 P~,
\label{qbox}
\end{equation}

\noindent
where now the operator $P$ projects onto the SM model space and
$Q=\mathbf{1}-P$.
In our calculations the $\hat{Q}$ box is expanded as a collection of
one- and two-body irreducible valence-linked Goldstone diagrams up to
third order in the perturbative expansion
\cite{Coraggio10a,Coraggio12a}.
We include intermediate states whose unperturbed
  excitation energy is less than$E_{max}=N_{max} \hbar \omega$, with
  $N_{max}=18$.
  This value of $N_{max}$ , as shown in \cite{Coraggio18b}, is large
  enough to obtain converged SP energy spacings and TBME.

Kuo and Krenciglowa have shown that the effective Hamiltonian can be
expressed as an infinite summation $H_{\rm eff} = \sum_{i=0}^{\infty}
F_i$, the terms $F_i$ being defined as a combination of $\hat{Q}$ box
and its derivatives \cite{Krenciglowa74}:

\begin{eqnarray}
F_0 & = & \hat{Q}(\epsilon_0)  \nonumber \\
F_1 & = & \hat{Q}_1(\epsilon_0)\hat{Q}(\epsilon_0)  \nonumber \\
F_2 & = & \left[ \hat{Q}_2(\epsilon_0)\hat{Q}(\epsilon_0) + 
\hat{Q}_1(\epsilon_0)\hat{Q}_1(\epsilon_0) \right] \hat{Q}(\epsilon_0)  \nonumber \\
~~ & ... & ~~ ,
\label{kkeqexp}
\end{eqnarray}
\noindent
where

\begin{equation}
\hat{Q}_m = \frac {1}{m!} \frac {d^m \hat{Q} (\epsilon)}{d \epsilon^m} \biggl| 
_{\epsilon=\epsilon_0} ~.
\label{qm}
\end{equation}

\noindent
The $Q$-box derivatives, that are energy-dependent, are calculated for
an energy value $\epsilon_0$ which corresponds to the model-space
eigenvalue of the unperturbed Hamiltonian $H_0$, that we have chosen
to be harmonic-oscillator (HO) one.
The HO parameters of the unperturbed Hamiltonians are
  $\hbar \omega=11,10,8$ MeV for model spaces placed above
  $^{40}$Ca, $^{56}$Ni, and $^{100}$Sn cores and according to the expression
  \cite{Blomqvist68} $\hbar \omega= 45 A^{-1/3} -25 A^{-2/3}$  for
  $A=40,56,100$, respectively.

$H_{\rm eff}$ provides the basic inputs of our SM calculations, namely
the single-particle (SP) energies, the two-body matrix elements (TBMEs)
of the residual interaction, and the calculated values that we have
employed for the diagonalization of the SM Hamiltonian are all
reported in Ref. \cite{Coraggio17a,Coraggio19a}, for the
  chosen model spaces.
In the same paper, we have discussed the perturbative behavior of
the calculated energy spectra, and a detailed discussion of the
perturbative properties of the derivation of $H_{\rm eff}$ can be also
found in Ref. \cite{Coraggio18b}.

\subsection{Effective two-body decay operators}
As mentioned in the introduction, the effective decay operators can be
constructed consistently with the derivation of \heff, using the
formalism presented by Suzuki and Okamoto in Ref. \cite{Suzuki95}, an
approach we have applied in Refs. \cite{Coraggio17a,Coraggio19a}.

As reported in the above mentioned papers, a non-Hermitian effective
\zbb -decay operator $\Theta_{\rm eff}$ can be expressed in terms of
\heff, the $\hat{Q}$ box and its derivatives, and an infinite sum of
operators $\chi_n$, as follows:

\begin{equation}
\Theta_{\rm eff}  = H_{\rm eff} \hat{Q}^{-1}  (\chi_0+ \chi_1 + \chi_2 +\cdots) ~.
\label{effopexp2}
\end{equation}

The $\chi_n$ operators are defined in terms of vertex functions
$\hat{\Theta} (\epsilon)$, $\hat{\Theta} (\epsilon_1, \epsilon_2)$ and
their derivatives, these vertex functions being constructed from a
bare \zbb -decay operator $\Theta$ analogously to the \qbox~definition:

\begin{eqnarray}
\hat{\Theta} (\epsilon) = & P \Theta P + P \Theta Q
\frac{1}{\epsilon - Q H Q} Q H_1 P ~, ~~~~~~~~~~~~~~~~~~~\label{thetabox} \\
\hat{\Theta} (\epsilon_1 ; \epsilon_2) = & P H_1 Q
\frac{1}{\epsilon_1 - Q H Q} Q \Theta Q \frac{1}{\epsilon_2 - Q H Q} Q H_1 P ~.~~~~~
\end{eqnarray}

Then, the $\chi_n$ operators \cite{Suzuki95} are written as:

\begin{eqnarray}
\chi_0 &=& (\hat{\Theta}_0 + h.c.)+ \hat{\Theta}_{00}~,  \label{chi0} \\
\chi_1 &=& (\hat{\Theta}_1\hat{Q} + h.c.) + (\hat{\Theta}_{01}\hat{Q}
+ h.c.) ~, \\
\chi_2 &=& (\hat{\Theta}_1\hat{Q}_1 \hat{Q}+ h.c.) +
(\hat{\Theta}_{2}\hat{Q}\hat{Q} + h.c.) + \nonumber \\
~ & ~ & (\hat{\Theta}_{02}\hat{Q}\hat{Q} + h.c.)+  \hat{Q}
\hat{\Theta}_{11} \hat{Q}~, \label{chin} \\
&~~~& \cdots \nonumber
\end{eqnarray}

\noindent
where $\hat{\Theta}_m$, $\hat{\Theta}_{mn}$ have the following
expressions:
\begin{eqnarray}
\hat{\Theta}_m & = & \frac {1}{m!} \frac {d^m \hat{\Theta}
 (\epsilon)}{d \epsilon^m} \biggl|_{\epsilon=\epsilon_0} ~, \\
\hat{\Theta}_{mn} & = & \frac {1}{m! n!} \frac{d^m}{d \epsilon_1^m}
\frac{d^n}{d \epsilon_2^n}  \hat{\Theta} (\epsilon_1 ;\epsilon_2)
\biggl|_{\epsilon_1= \epsilon_0, \epsilon_2  = \epsilon_0} ~.
\end{eqnarray}

As in our previous work \cite{Coraggio19a}, where we have calculated
effective electromagnetic transition and $\beta$-decay operators, we
have derived $\Theta_{\rm eff}$, which accounts for the truncation to
the reduced SM space, arresting the $\chi_n$ series to the $\chi_2$
term.
It is worth pointing out that $\chi_3$ depends on the first, second,
and third derivatives of $\hat{\Theta}_0$ and $\hat{\Theta}_{00}$, and
on the first and second derivatives of the $\hat{Q}$ box (see
Eq. (\ref{chin})), so we estimate $\chi_3$ contribution being at least
one order of magnitude smaller than the $\chi_2$ one. 

\begin{figure}[ht]
\begin{center}
\includegraphics[scale=0.60,angle=0]{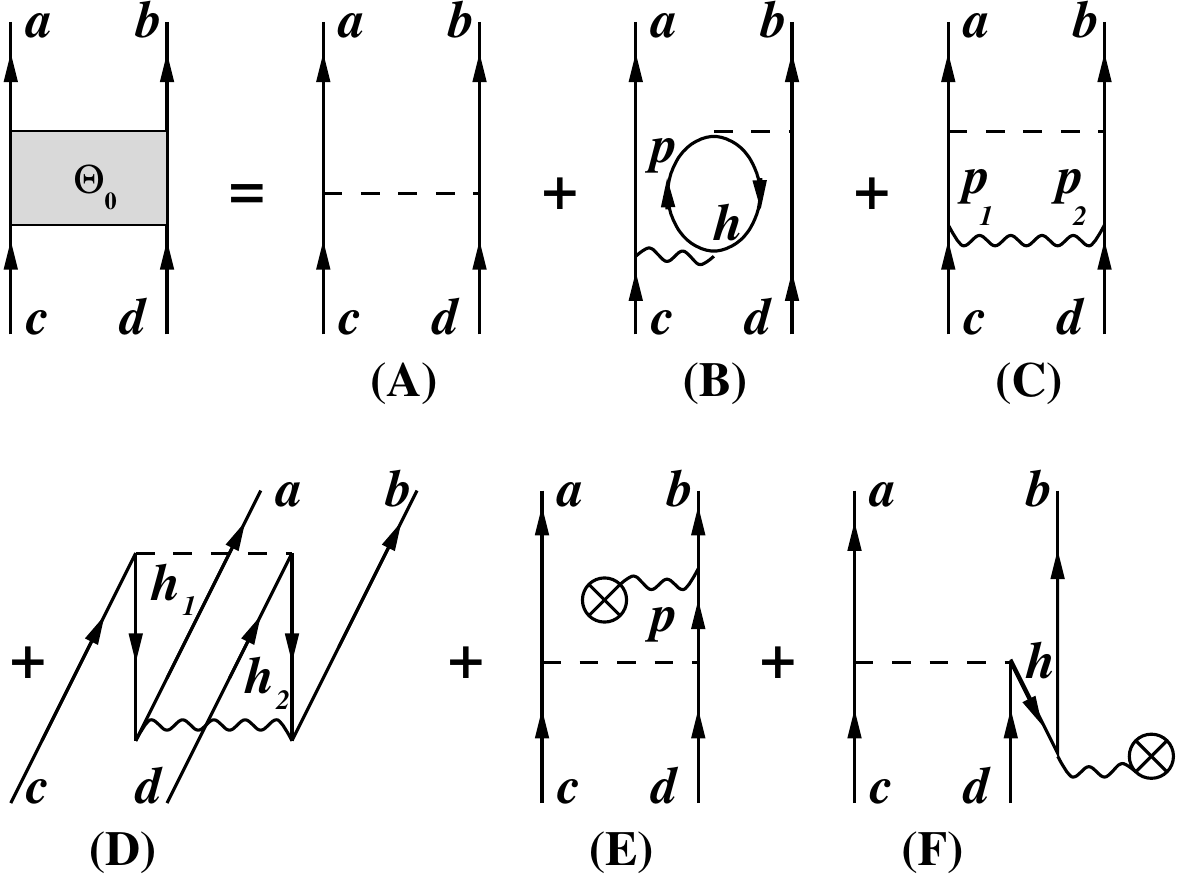}
\caption{Second-order diagrams included in the perturbative
  expansion of $\hat{\Theta}$. The dashed lines indicate the bare operator
  $\Theta$, the wavy lines the two-body potential $V_\lwk$ (see
  text for details).}
\label{figeffop2b}
\end{center}
\end{figure}

The calculation of $\chi_0$, $\chi_1$, and $\chi_2$ is performed
carrying out a perturbative expansion of $\hat{\Theta}_0$ and
$\hat{\Theta}_{00}$, including diagrams up to the third order in the
perturbation theory, consistently with the perturbative expansion of
the \qbox.
In Fig. \ref{figeffop2b} we report all the two-body $\Theta_0$
diagrams up to the second order, the bare operator $\Theta$ being
represented with a dashed line.
The first-order $(V_\lwk-U)$-insertion, represented by a
circle with a cross inside, arises because of the presence of the $-U$
term in the interaction Hamiltonian $H_1$ (see for example
Ref. \cite{Coraggio12a} for details).

We point out that diagrams (A)-(D) belongs also to the perturbative
expansion of $\Theta_{\rm eff}$ in Refs. \cite{Wu85,Holt13d}, where
diagrams (E) and (F) were neglected.
The contribution of $(V_\lwk-U)$-insertion diagrams, such as (E) and
(F), is equal to zero only under the hypothesis that the HO potential
would correspond to a Hartree-Fock basis for the \vlwk~potential.
In a previous study, we have shown that the role of this class of
diagrams is non-negligible to derive \heff, in particular to benchmark
RSM with {\it ab initio} calculations \cite{Coraggio12a}.

So far we have presented the derivation of an effective operator just
for a nuclear system with two valence nucleons, but in the following
section we are going to focus on \zbb~decay of nuclei that, within the
shell-model, will be described in terms of a number of valence
nucleons that is much larger than 2.
For example double-$\beta$ decay of $^{136}$Xe into $^{136}$Te
involves 36 valence nucleons outside the doubly-magic $^{100}$Sn, and
in such a case the expression of $\Theta_{\rm eff}$ should contain
contribution up to a 36-body term.

\begin{figure}[H]
\begin{center}
\includegraphics[scale=0.60,angle=0]{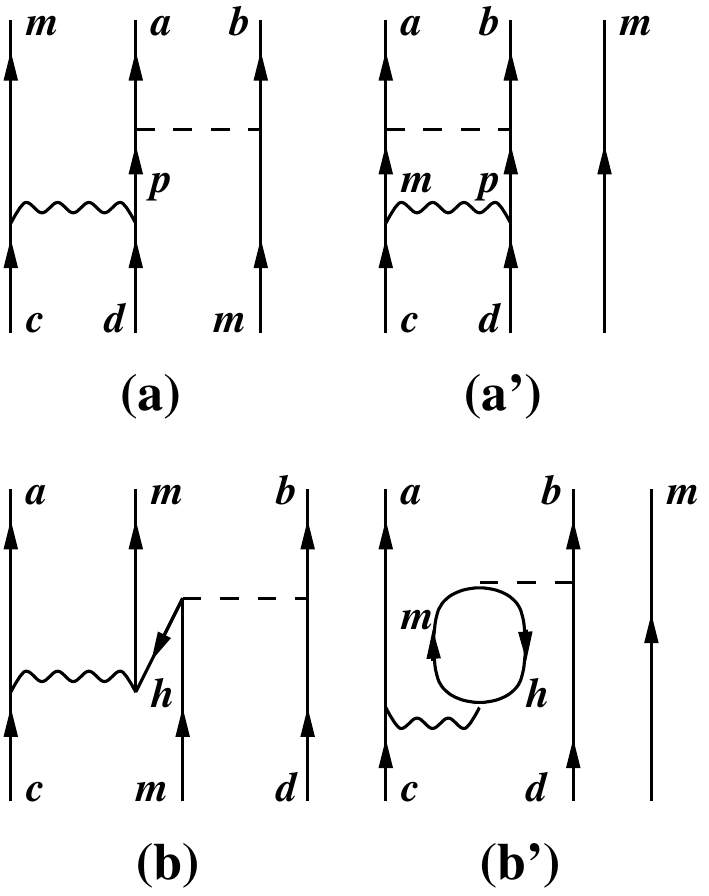}
\caption{Second-order three-body diagrams which are included in the
  perturbative expansion of $\hat{\Theta}$.
  As in Fig. \ref{figeffop2b}, the dashed line indicates the bare operator 
  $\Theta$, the wavy line the two-body potential $V_\lwk$.}
\label{figeffop3b}
\end{center}
\end{figure}

At present this is unfeasible, so we include just the leading terms of
these many-body contributions in the perturbative expansion of
$\hat{\Theta}$, namely the second-order three-body diagrams (a)
and (b), that are reported in Fig. \ref{figeffop3b}.
For the sake of simplicity, for each topology only one of the diagrams
which correspond to the permutation of the external lines is drawn.
  
The two topologies of second-order connected three-valence-nucleon
diagrams (a) and (b) correct the  Pauli-principle violation introduced by
diagram (a') and (b') when one of the intermediate
particle states is equal to $m$ \cite{Ellis77}.
This is the so-called ``blocking effect'', which urges to take into
account the Pauli exclusion principle in systems with more than two
valence nucleons \cite{Towner87}.

It should be pointed out that also the authors of Ref. \cite{Holt13d} have
attempted to account for this effect in an approximate way, by weighting
the intermediate model-space-particle lines that appear in the
$\hat{\Theta}$-box diagrams with a factor that suppress their matrix
elements in terms of the unperturbed occupation density of the line
orbital.

Since the NATHAN SM code, that we employ to calculate
\nme~\cite{NATHAN}, cannot manage three-body decay operators, we have
derived a density-dependent two-body contribution at one-loop order
from the three-body diagrams in Fig. \ref{figeffop3b}, summing over
the partially-filled model-space orbitals.
The details of this procedure can be found in Ref. \cite{Itaco19}, as
well as in Ref. \cite{Ma19}, where the same procedure has been
preformed to derive density-dependent two-body \heffs~ to study
many-valence nucleon systems.

\subsection{The \zbb-decay operator}
We now focus our attention on the vertices of the bare \zbb~operator
$\Theta$.

We recall that the formal expression of $M_{\alpha}^{0\nu}$ - $\alpha$
denoting the Fermi ($F$), Gamow-Teller (GT), or tensor ($T$) decay
channels - is written in terms of the one-body transition-density
matrix elements between the daughter and parent nuclei (grand-daughter
and daughter nuclei) $ \langle k | a^{\dagger}_{p^\prime} a_{n^\prime}
| i \rangle$ ($ \langle f | a^{\dagger}_{p}a_{n} | k \rangle $), $p,n$
subscripts denoting proton and neutron states, and indices $i,k,f$ the
parent, daughter, and grand-daughter nuclei, respectively:

\begin{eqnarray}
 M_\alpha^{0\nu} & =  & \sum_{k} \sum_{j_p j_{p^\prime} j_n
                        j_{n^\prime} } \langle f | a^{\dagger}_{p}a_{n} | k \rangle \langle k |
 a^{\dagger}_{p^\prime} a_{n^\prime} | i \rangle \nonumber \\
~ & ~& \times \left< j_p  j_{p^\prime} \mid \tau^-_{1}
\tau^-_{2} \Theta_\alpha^{k} \mid j_n j_{n^\prime} \right> ~. \label{M0nu}
\end{eqnarray}

The operators $\Theta_{\alpha}$ are expressed in terms of the
neutrino potentials $H_{\alpha}$ and form functions $h_{\alpha}(q)$:
\begin{eqnarray} 
 \Theta^{k}_{\rm GT} & = & \vec{\sigma}_1 \cdot \vec{\sigma}_2 H_{\rm
GT}^k(r) \label{operatorGT} \\
\Theta^{k}_{\rm F} & = & H_{\rm F}^k(r)  \label{operatorF} \\ 
\Theta^{k}_{\rm T} & = & \left[3\left(\vec{\sigma}_1 \cdot \hat{r} \right) 
\left(\vec{\sigma}_1 \cdot \hat{r} \right) - \vec{\sigma}_1 \cdot
                      \vec{\sigma}_2 \right] H_{\rm T}^k(r)~, \label{operatorT}
\end{eqnarray}

\begin{equation}
H_{\alpha}^k(r)=\frac {2R}{\pi} \int_{0}^{\infty} \frac {j_{n_{\alpha}}(qr)
  h_{\alpha}(q^2)qdq}{q+E_k-(E_i+E_f)/2}~.
\label{neutpot}
\end{equation}

The value of the parameter $R$ is $R=1.2 A^{1/3}$ fm, the
$j_{n_{\alpha}}(qr)$ are the spherical Bessel functions,
$n_{\alpha}=0$ for Fermi and Gamow-Teller components, $n_{\alpha}=2$
for the tensor one.
For the sake of clarity, the explicit expression of neutrino form
functions $h_{\alpha}(q)$ for light-neutrino exchange we employ in
present calculations are reported below:

\begin{eqnarray}
h_{F} ({ q}^{2})  & = & g^2_V({ q}^{2}) \nonumber \\
h_{GT} ({ q}^{2}) & = & \frac{g^2_A({ q}^{2})}{g^2_A} 
\left[ 1 - \frac{2}{3} \frac{ { q}^{2}}{ { q}^{2} + m^2_\pi } + 
\frac{1}{3} ( \frac{ { q}^{2}}{ { q}^{2} + m^2_\pi } )^2 \right]
\nonumber\\
&& + \frac{2}{3} \frac{g^2_M({ q}^{2} )}{g^2_A} \frac{{ q}^{2} }{4 m^2_p }, 
\nonumber \\
h_T ({ q}^{2}) & = & \frac{g^2_A({ q}^{2})}{g^2_A} \left[ 
\frac{2}{3} \frac{ { q}^{2}}{ { q}^{2} + m^2_\pi } -
\frac{1}{3} ( \frac{ { q}^{2}}{ { q}^{2} + m^2_\pi } )^2 \right]
\nonumber\\
&& + \frac{1}{3} \frac{g^2_M ({ q}^{2} )}{g^2_A} \frac{{ q}^{2} }{4 m^2_p }    
\end{eqnarray}

The dipole approximation has been adopted for the vector, axial-vector
and weak-magnetism $g_V({ q}^{2}), g_A({ q}^{2}), g_M({ q}^{2})$ form
factors :
\begin{eqnarray}
g_V({ q}^{2})&=& \frac{g_V}{(1+{ q}^{2}/{\Lambda^2_V})^2}, 
\nonumber\\
g_M({ q}^{2}) &=& (\mu_p-\mu_n) g_V({ q}^{2}), 
\nonumber\\
g_A({ q}^{2}) &=& \frac{g_A}{(1+{ q}^{2}/{\Lambda^2_A})^2},
\end{eqnarray}
where $g_V = 1$, $g_A \equiv g_A^{free}=1.2723$,  $(\mu_p - \mu_n) =
3.70$, and the cutoff parameters $\Lambda_V = 850$ MeV and $\Lambda_A
= 1086$ MeV.

The expression in Eq. (\ref{M0nu}) can be easier managed
computationally within the QRPA, while all other models - including
most of SM calculations -  resort to the so-called closure
approximation.
This approximation is based on the fact that the relative momenta $q$
of the neutrinos involved in the decay, appearing in the denominator
of the neutrino potential of Eq. (\ref{neutpot}), are of the order of
100-200 MeV, while the excitation energies of the nuclei involved in
the \zbb decay are only of the order of 10 MeV \cite{Senkov13}.

The above considerations lead to replace the energies of the
intermediate states $E_k$ in Eq. (\ref{neutpot}) by an average value
$E_k-(E_i+E_f)/2 \rightarrow \langle E \rangle$, and simplify both the
relationships (\ref{M0nu}, \ref{neutpot}).
$M_\alpha^{0\nu}$ can be re-written in terms of the
two-body transition-density matrix elements $\langle f |
a^{\dagger}_{p}a_{n} a^{\dagger}_{p^\prime} a_{n^\prime} | i \rangle$:

\begin{eqnarray}
M_\alpha^{0\nu}& =  & \sum_{j_n j_{n^\prime} j_p j_{p^\prime}}
\langle f | a^{\dagger}_{p}a_{n} a^{\dagger}_{p^\prime} a_{n^\prime} 
| i \rangle \nonumber \\
~ & ~& \times  \left< j_p  j_{p^\prime} \mid \tau^-_{1}
\tau^-_{2} \Theta_\alpha \mid  j_n j_{n^\prime}
       \right>~, \label{M0nuapp}
\end{eqnarray}
 
\noindent
and the neutrino potential are expressed as:
\begin{equation}
H_{\alpha}(r)=\frac {2R}{\pi} \int_{0}^{\infty} \frac {j_{n_{\alpha}}(qr)
  h_{\alpha}(q^2)qdq}{q+\left< E \right>}~.
\label{neutpotapp}
\end{equation}

\noindent
In present calculations, we adopt the closure approximation to define
the operators $\Theta$ in Eqs. (\ref{operatorGT}-\ref{operatorT}), and
the average energies $\left< E \right>$ have been evaluated as in
Refs. \cite{Haxton84,Tomoda91}.  
It should be noted that the closure approximation simplifies the
derivation of $\Theta_{\rm eff}$, since the diagrams appearing in the
perturbative expansion of the vertex function $\hat{\Theta}$ will not
be dependent on the energy of the daughter-nucleus intermediate states
$E_k$.

In this regard, the authors of Ref. \cite{Senkov13} have
  performed SM calculations of \nme~for $^{48}$Ca \zbb~decay both
  within and beyond closure approximation, and have shown that the
  results beyond closure-approximation are $10\%$ larger.

Now, it is worth to recollect the main motivations to renormalize the
\zbb-decay operator, as we have already mentioned in the Introduction:
\begin{enumerate}
\item[a)] The truncation of the $A$-body problem in the full Hilbert
  space to the problem of few valence nucleons interacting in the
  reduced model space.
\item[b)] The contribution of the short-range correlations (SRC),
  accounting the fact that the action of a two-body decay operator on
  an unperturbed (uncorrelated) wave function, that is used in the
  perturbative expansion of $\Theta_{\rm eff}$ is not equal to the
  action of the same operator on the real (correlated) nuclear wave
  function.
\item[c)] The contribution of the two-body meson-exchange corrections
  to the electroweak currents, originated from sub-nucleonic degrees
  of freedom.
\end{enumerate}

Up to this point, we have extensively covered point (a), and we need
to discuss about issues (b) and (c).

As regards the inclusion of the SRC, in Ref. \cite{Coraggio19b} we
have introduced an original approach that is consistent with the
\vlwk~procedure.
More precisely, the \zbb~ operator $\Theta$ is calculated within the
momentum space and then renormalized  by way of $\Omega$ in order to
consider effectively the high-momentum (short range) components of the
$NN$ potential, in a framework where their direct contribution is
dumped by the introduction of a cutoff $\Lambda$.

Consequently, the $\Theta$ vertices appearing in the perturbative
expansion of the $\hat{\Theta}$ box are subsituted with the renormalized
$\Theta_\lwk$ operator that is defined as $\Theta_\lwk \equiv P \Omega
\Theta \Omega^{-1} P$ for relative momenta $k < \Lambda$, and is equal
to zero for $k > \Lambda$.

We have found that the effect in magnitude of this renormalization
procedure is similar to the SRC modeled by the so-called Unitary
Correlation Operator Method (UCOM) \cite{Menendez09b}, meaning a
lighter softening of \nme~with respect to the one provided by
Jastrow type SRC. 

The last issue to be considered is the the role of meson-exchange
corrections to the electroweak currents, that has been also
investigated in Refs. \cite{Menendez11,Wang18}.
The framework of these studies is the Chiral Effective Field Theory
(ChEFT) that allows a consistent derivation of nuclear two- and
three-body forces, as well as electroweak chiral two-body currents
which are characterized by the same low-energy constants (LECs)
appearing in the structure of the nuclear Hamiltonian.

In our calculations we start from the CD-Bonn potential, and then
renormalize it by way of the \vlwk~ procedure, preventing us to
approach issue (c) in a way that is consistent with the derivation of
our starting two-body force (as we do for the treatment of SRC).
However, we are confident that, since we employ a large cutoff
($\Lambda=2.6$ fm$^{-1}$) for the derivation of  \vlwk, our results
are less affected by residual three-body force contributions and,
consequently, by electroweak two-body current corrections.

\section{Results}\label{calculations}

\begin{figure}[h]
\begin{center}
\includegraphics[scale=0.47,angle=0]{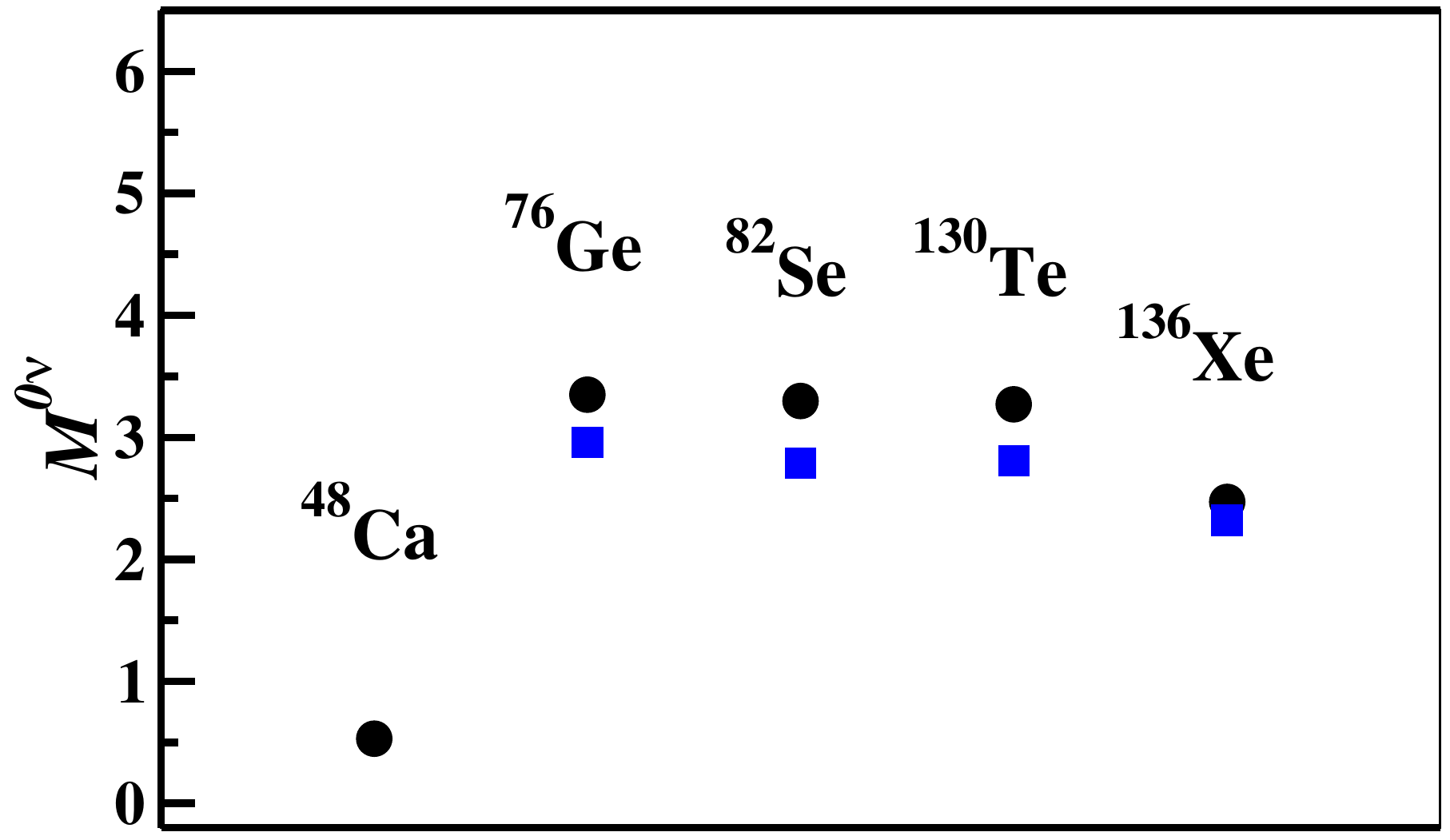}
\caption{Calculated values of \nme~for $^{48}$Ca, $^{76}$Ge, $^{82}$Se,
$^{130}$Te, and $^{136}$Xe decays (black dots), obtained from the bare
\zbb~decay operator. The results are compared with those reported in
Ref. \cite{Menendez09b} (blue squares), and obtained without including
SRC.}
\label{figSMcomparison}
\end{center}
\end{figure}

We recall that our calculations of \nme~account for the light-neutrino
exchange mechanism, and that the effective Hamiltonians we have
employed are those reported in Refs.\cite{Coraggio17a,Coraggio19a},
where it can be found a detailed description of the low-energy
spectroscopic properties of $^{48}$Ca, $^{76}$Ge, $^{82}$Se,
$^{130}$Te, and $^{136}$Xe, obtained with their diagonalization.
In present work we neglect the contribution of the tensor component of
Eq. (\ref{operatorT}), since it plays a minor role, its matrix elements
being about two order of magnitude smaller than those corresponding to
the Gamow-Teller and Fermi decay operators.
Actually, it should be pointed out that the $^{48}$Ca
  decay is an exception, since the ratio of the tensor component to
  the GT contribution may reach the value of $10\%$ \cite{Menendez09b}.

In our calculations, the total nuclear matrix element $M^{0\nu}$ is
expressed as

\begin{equation}
 M^{0 \nu} = M^{0 \nu}_{\rm GT}- \left( \frac{g_{V}}{g_{A}} \right)^2  M^{0
   \nu}_{\rm F}~,
\label{nme00nu}
\end{equation}
\noindent
and depends on the axial and vector coupling constants $g_A,g_V$, the
free values being $g_A^{free}=1.2723, ~ g_V^{free}=1$ \cite{PDG18}.

The nuclear matrix elements $M^{0\nu}_{\rm GT,F}$ are calculated
accordingly
Eqs. (\ref{operatorGT},\ref{operatorF},\ref{M0nuapp},\ref{neutpotapp}),
namely within the closure approximation. 

The first results it is worth presenting are the calculated \nme
values for the nuclei under investigation, obtained with the bare
decay-operator $\Theta$, without any contribution from the
renormalization procedure (see Fig. \ref{figSMcomparison}).

We compare the results of our calculations (black dots) with those
obtained in similar SM calculations employing the same model spaces
and the same \zbb~decay operator, that is the one without any kind of
renormalization.
More precisely, we refer to the calculations by Madrid-Strasbourg
group (blue squares) (see the first column in Table 8 in
  Ref. \cite{Menendez09b}).
It is worth pointing out that the results represented by the blue
squares include also the tensor component of \nme, but the authors
have found that $M^{0 \nu}_{\rm T}$ contribution for $^{76}$Ge,
$^{82}$Se, $^{130}$Te, and $^{136}$Xe decays is less than $1\%$ of the
dominant GT one \cite{Menendez09b}.

Since the two-body matrix elements of $\Theta$ are the same, this
comparison can be considered a sort of benchmark of the nuclear SM
wave functions, that have been obtained starting from different
\heffs.
As a matter of fact, the TBME of \heffs~employed in
Ref. \cite{Menendez09b} have been derived as a fine tuning of TBME
obtained from realistic SM \heffs~\cite{Hjorth95}.
The modifications have been made to fit to some specific spectroscopic
data of nuclei in $pf$, $f_{5/2}pg_{9/2}$, and $g_{7/2}dsh_{11/2}$
regions, as well as single-particle properties in the same regions to
determine the SP energies (details can be found in the above mentioned
papers and references therein).

\begin{figure}[h]
\begin{center}
\includegraphics[scale=0.32,angle=0]{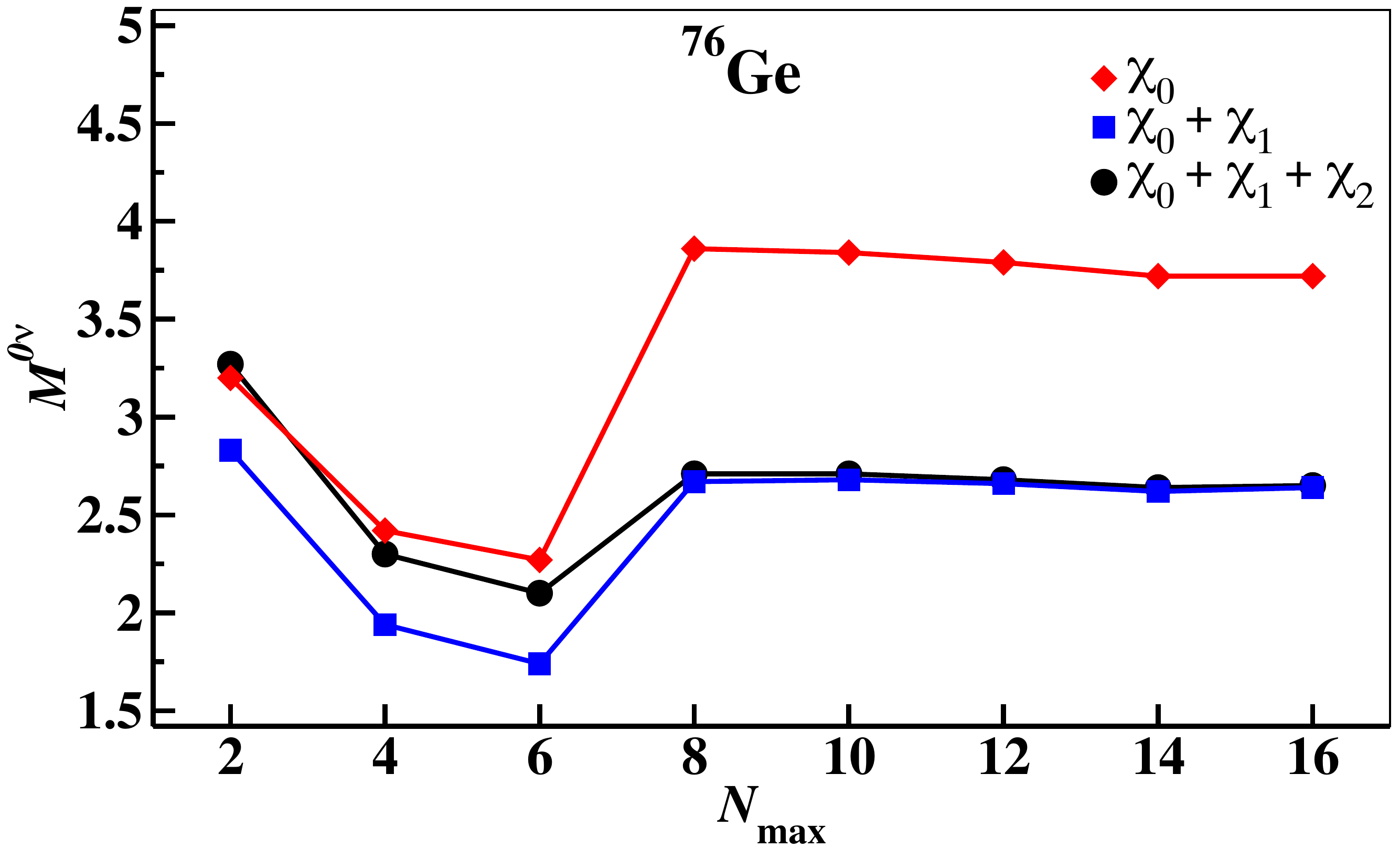}
\caption{\nme~for the $^{76}\mbox{Ge} \rightarrow ^{76}$Se decay as a
  function of $N_{\rm max}$. The red diamonds correspond to a
  truncation of $\chi_n$ expansion up to $\chi_0$, blue squares up to
  $\chi_1$, and black dots up to $\chi_2$.}
\label{figNmax}
\end{center}
\end{figure}

As described in Section \ref{outline}, our \heffs~have been derived
from a realistic $V_{NN}$ and their matrix elements have been not
modified to improve the agreement with experiment.
As can be seen from inspection of Fig. \ref{figSMcomparison}, there is
a general agreement among the different calculations, that is linked
to a common quality of the different \heffs~to reproduce
satisfactorily a large amount of spectra in these mass regions.

\begin{figure}[h]
\begin{center}
\includegraphics[scale=0.32,angle=0]{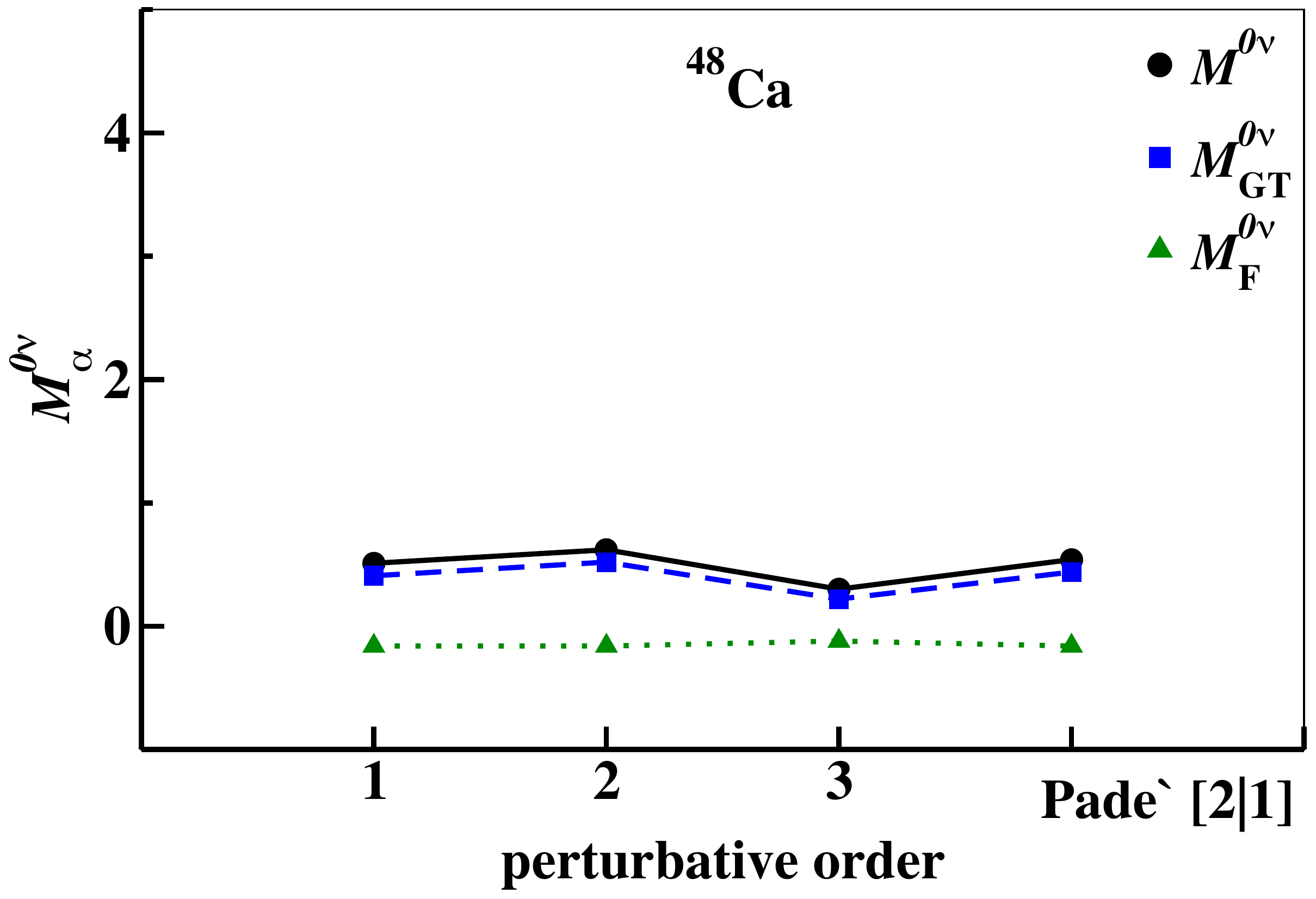}
\caption{\nme~for the $^{48}\mbox{Ca} \rightarrow ^{48}$Ti decay as a
  function of the perturbative order. The green triangles correspond
  to $M^{0\nu}_{\rm F}$, the blue squares to $M^{0\nu}_{\rm GT}$, and
  the black dots to the full \nme.}
\label{48Ca_obo}
\end{center}
\end{figure}

\begin{figure}[h]
\begin{center}
\includegraphics[scale=0.32,angle=0]{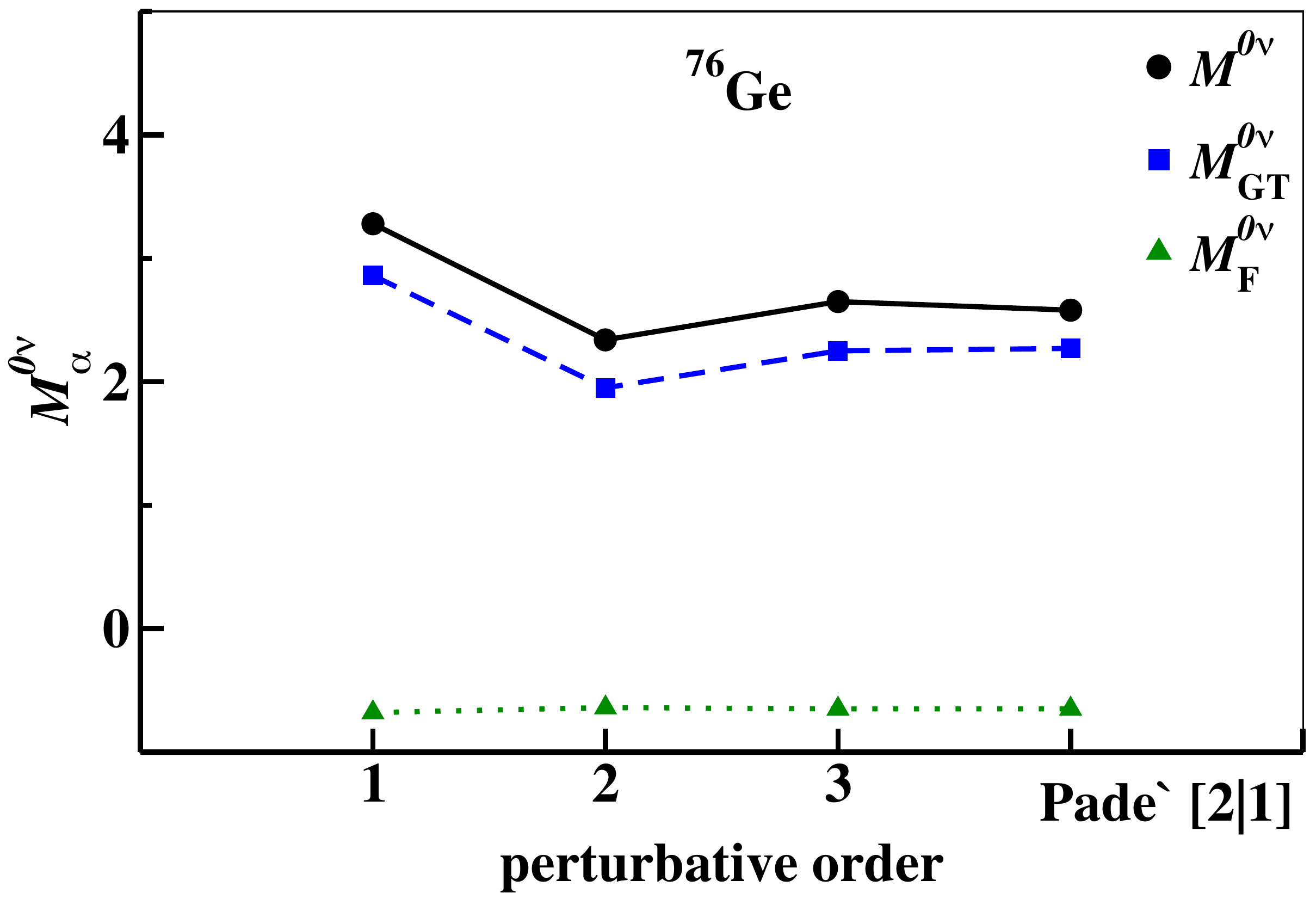}
\caption{ Same as in Fig. \ref{48Ca_obo}, but for the $^{76}\mbox{Ge}
  \rightarrow ^{76}$Se decay.}
\label{76Ge_obo}
\end{center}
\end{figure}

We shift now the focus on the results of the calculations obtained by
employing the effective decay-operator $\Theta_{\rm eff}$, which
accounts for the truncation of the Hilbert space, the SRC, and the
Pauli-blocking effect as well.

First, we report about the convergence properties with respect to the
number of intermediate states included in the perturbative expansion
of $\Theta_{\rm eff}$, and the truncation of the order of $\chi_n$
operators.

In Fig. \ref{figNmax} the calculated values of \nme~for the ${\rm
  ^{76}Ge} \rightarrow {\rm ^{76}Se}$ decay are reported as a function
of the maximum allowed excitation energy of the intermediate states
expressed in terms of the oscillator quanta $N_{\rm max}$, and for
contributions up to the $\chi_2$ operator.
The plot shows that the results are substantially convergent from
$N_{\rm max}=12$ on and the contributions from  $\chi_1$ are crucial
while those from $\chi_2$ are almost negligible.

It is worth pointing out that $\chi_3$ depends on the first, second,
and third derivatives of $\hat{\Theta}_0$ and $\hat{\Theta}_{00}$, as
well as on the first and second derivatives of the $\hat{Q}$ box (see
Eq. (\ref{chin})), so we estimate $\chi_3$ contribution being
at least one order of magnitude smaller than the $\chi_2$ one.

On the basis of the above analysis, the results we report in this
Section are all obtained including in the perturbative expansion up to
third-order diagrams, whose number of intermediate states corresponds
to oscillator quanta up to $N_{\rm  max}=14$, and up to $\chi_2$
contributions.

\begin{figure}[h]
\begin{center}
\includegraphics[scale=0.32,angle=0]{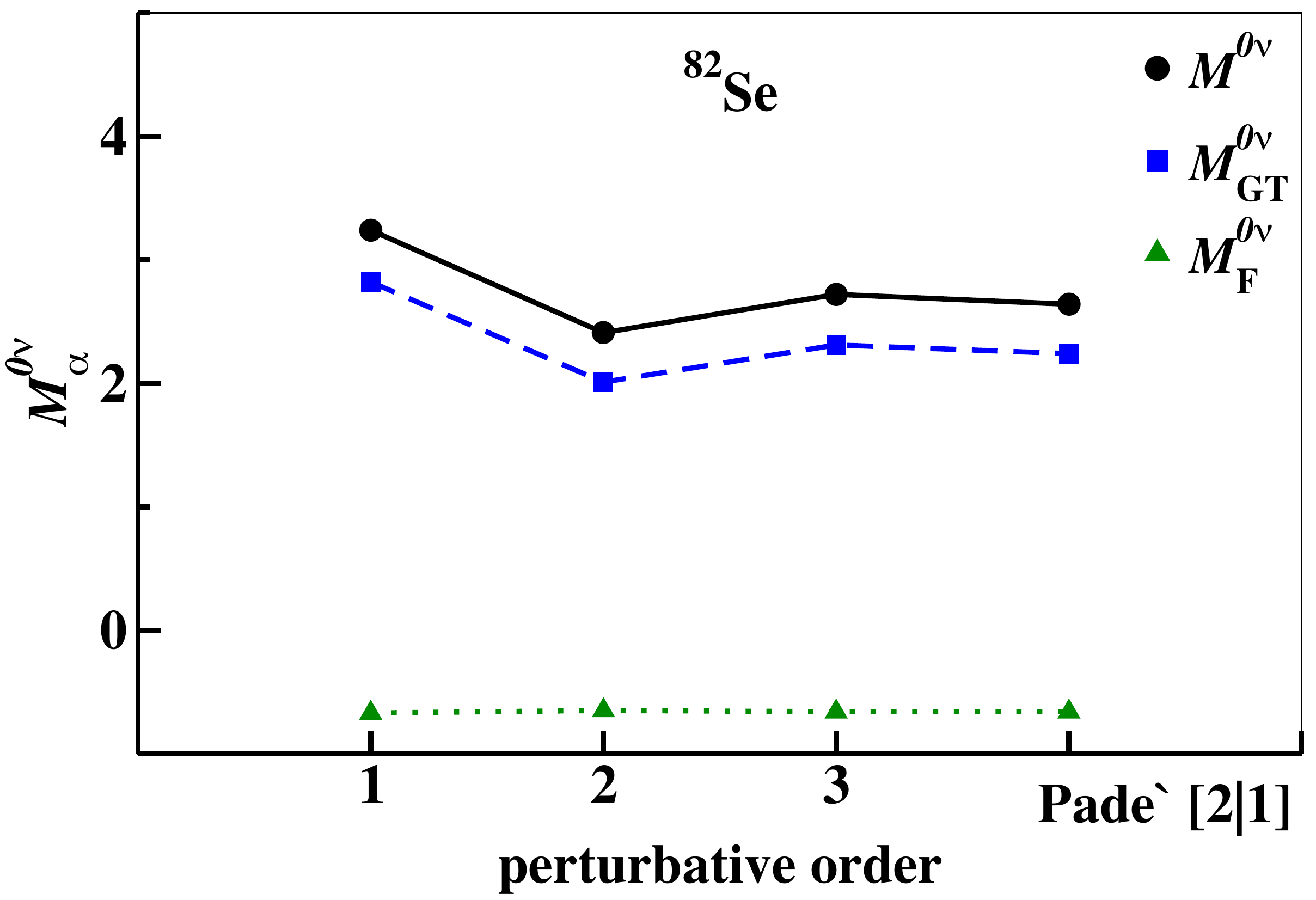}
\caption{ Same as in Fig. \ref{48Ca_obo}, but for the $^{82}\mbox{Se}
  \rightarrow ^{82}$Kr decay.}
\label{82Se_obo}
\end{center}
\end{figure}

\begin{figure}[h]
\begin{center}
\includegraphics[scale=0.32,angle=0]{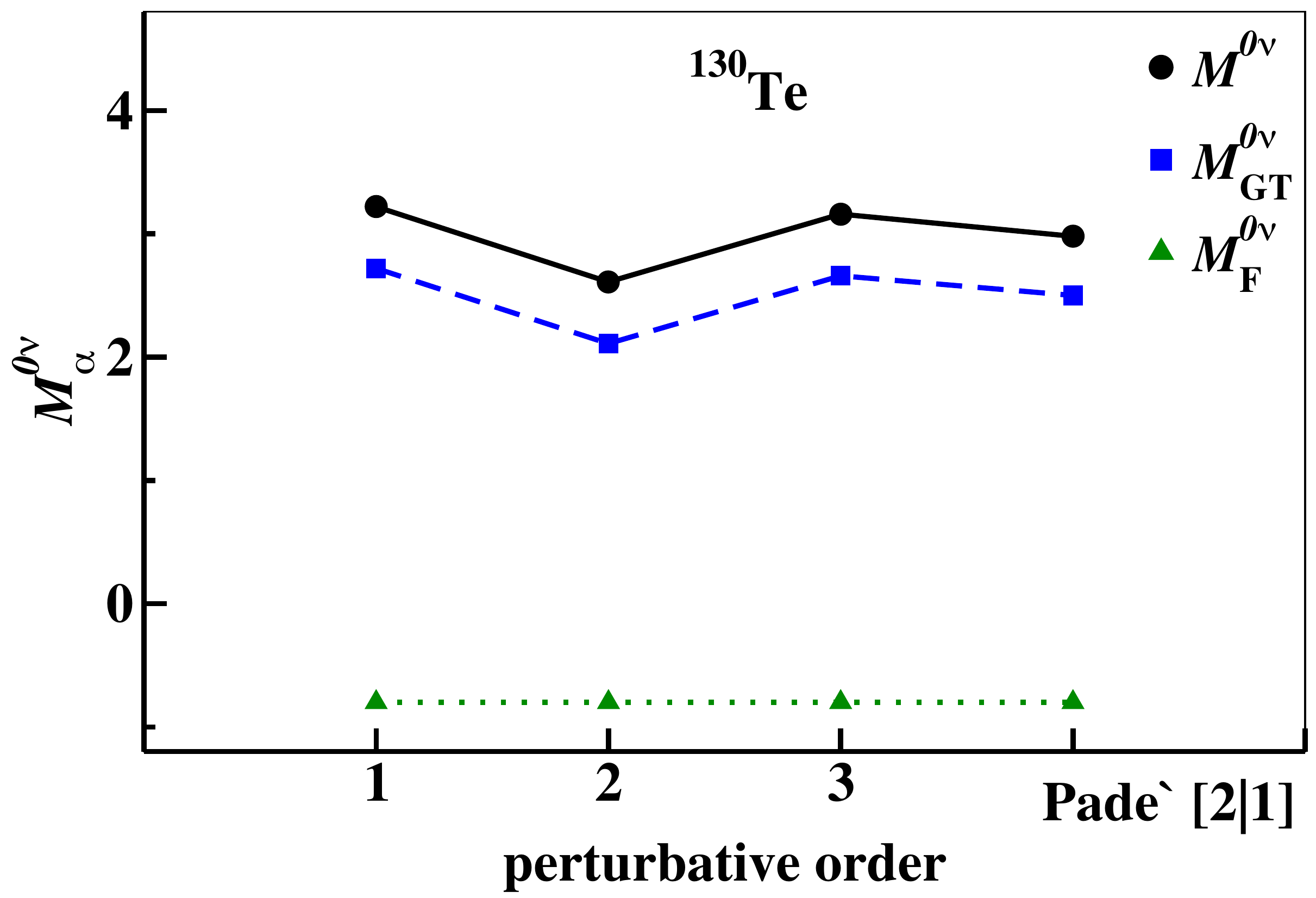}
\caption{ Same as in Fig. \ref{48Ca_obo}, but for the $^{130}\mbox{Te}
  \rightarrow ^{130}$Xe decay.}
\label{130Te_obo}
\end{center}
\end{figure}

\begin{figure}[h]
\begin{center}
\includegraphics[scale=0.32,angle=0]{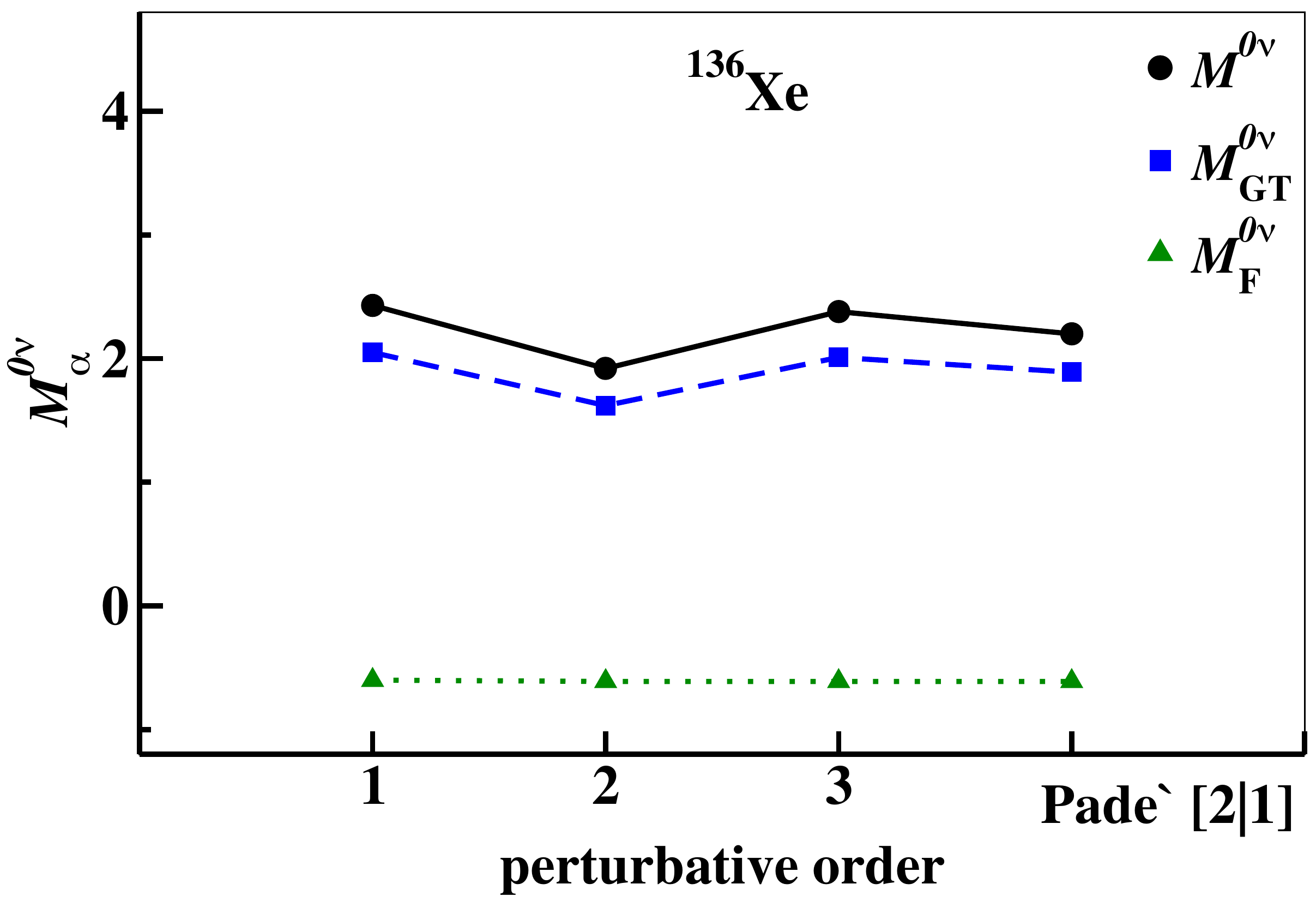}
\caption{ Same as in Fig. \ref{48Ca_obo}, but for the $^{136}\mbox{Xe}
  \rightarrow ^{136}$Ba decay.}
\label{136Xe_obo}
\end{center}
\end{figure}

Now, we consider the order-by-order convergence behavior by reporting
in Figs. \ref{48Ca_obo}-\ref{136Xe_obo} the calculated values of \nme,
$M^{0\nu}_{\rm GT}$, and $M^{0\nu}_{\rm F}$ for $^{48}$Ca, $^{76}$Ge,
$^{82}$Se, $^{130}$Te, and $^{136}$Xe\zbb~decay, respectively,  from
first- up to third-order in perturbation theory.
We compare the order-by-order results also with their Pad\'e
approximant $[2|1]$, as an indicator of the quality of
the perturbative behavior \cite{Baker70}.
It should be pointed out that the same scale has been employed in all
the figures.

First of all, we observe that the perturbative behavior is dominated
by the Gamow-Teller component, since the renormalization procedure
does not affect significantly the Fermi matrix element $M^{0\nu}_{\rm
  F}$.
We recall that the perturbative behavior of the single $\beta$-decay
operator provides a difference between the \nmed values calculated at
second and third order in perturbation theory which does not exceed
$10\%$ \cite{Coraggio18b}.
Here, we observe a less satisfactory perturbative behavior for our
calculation of \nme, the difference between second- and third-order
results being about $15\%$ and $30\%$ for $^{76}$Ge, $^{82}$Se, and
$^{130}$Te,$^{136}$Xe \zbb~decays, respectively.

\begin{figure}[h]
\begin{center}
\includegraphics[scale=0.32,angle=0]{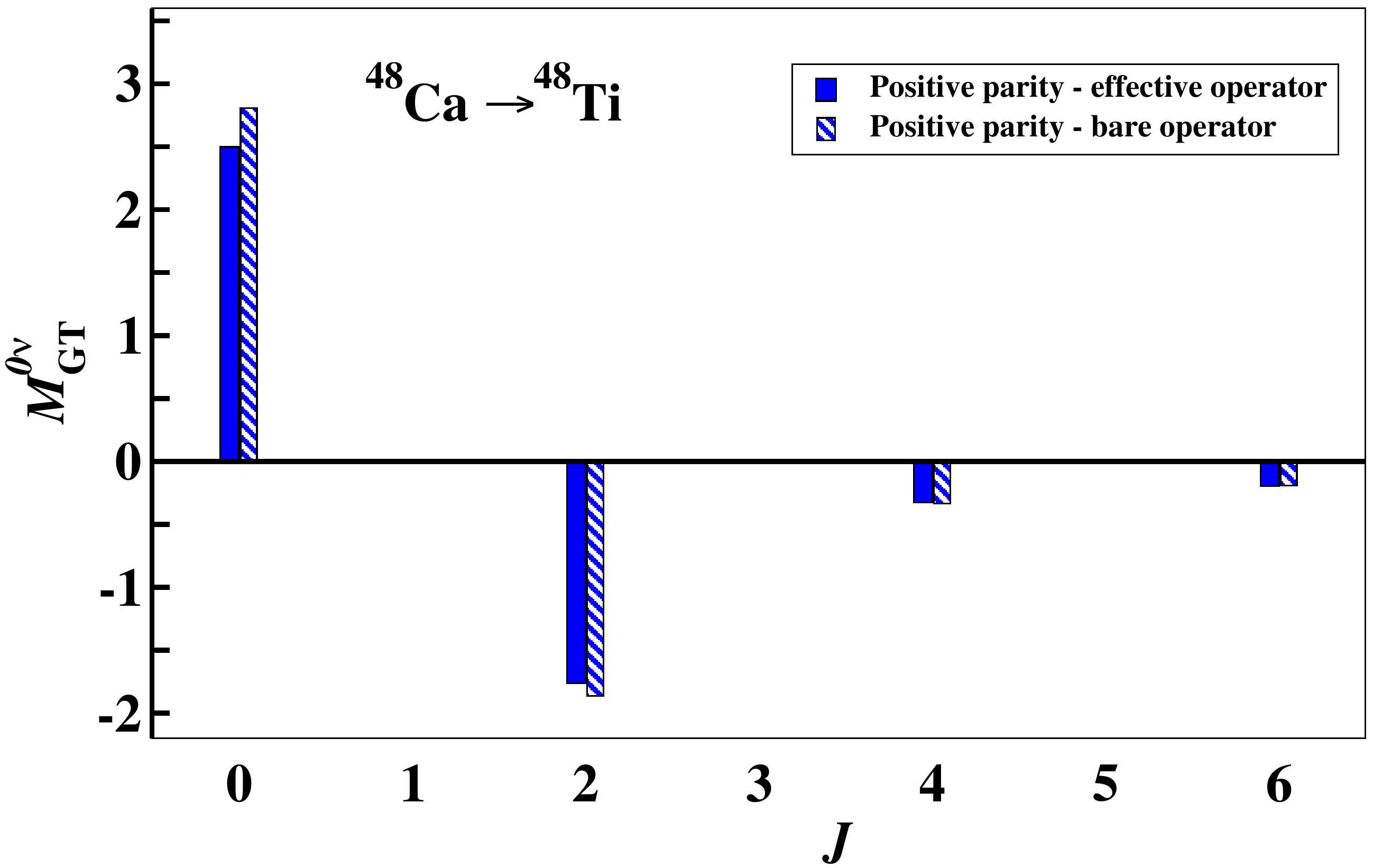}
\caption{Contributions from pairs of decaying neutrons with given
  $J^{\pi}$ to $M^{0\nu}_{\rm GT}$ for $^{48}$Ca \zbb~decay. The bars
  filled in blue corresponds to the results obtained with $\Theta_{\rm
    eff}$, those in dashed blue to the ones calculated with bare
  operator}
\label{48Ca-jj}
\end{center}
\end{figure}

\begin{figure}[h]
\begin{center}
\includegraphics[scale=0.32,angle=0]{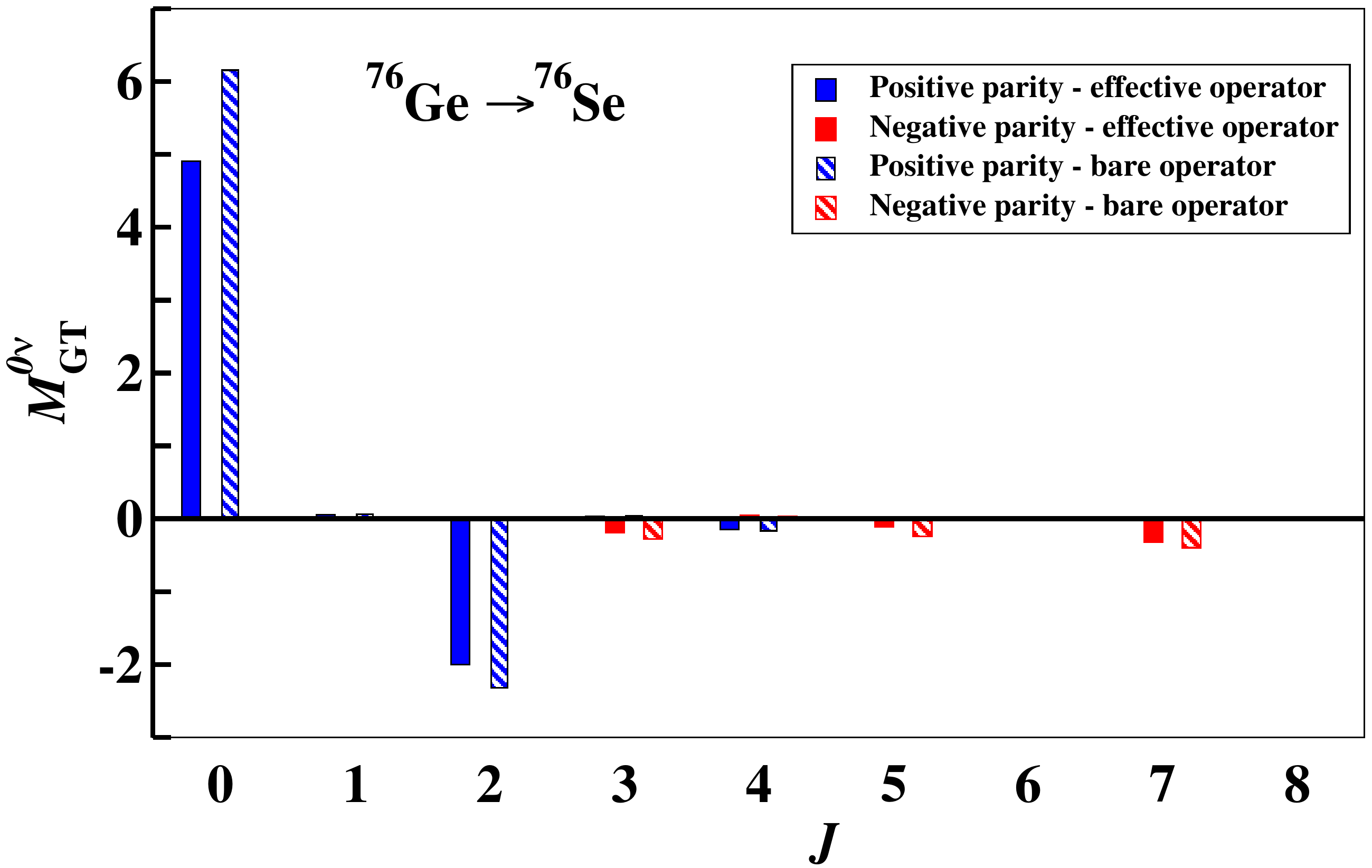}
\caption{Same as in Fig. \ref{48Ca-jj}, but for $^{76}$Ge
  \zbb~decay. The red bars corresponds to negative-parity states.}
\label{76Ge-jj}
\end{center}
\end{figure}

\begin{figure}[h]
\begin{center}
\includegraphics[scale=0.32,angle=0]{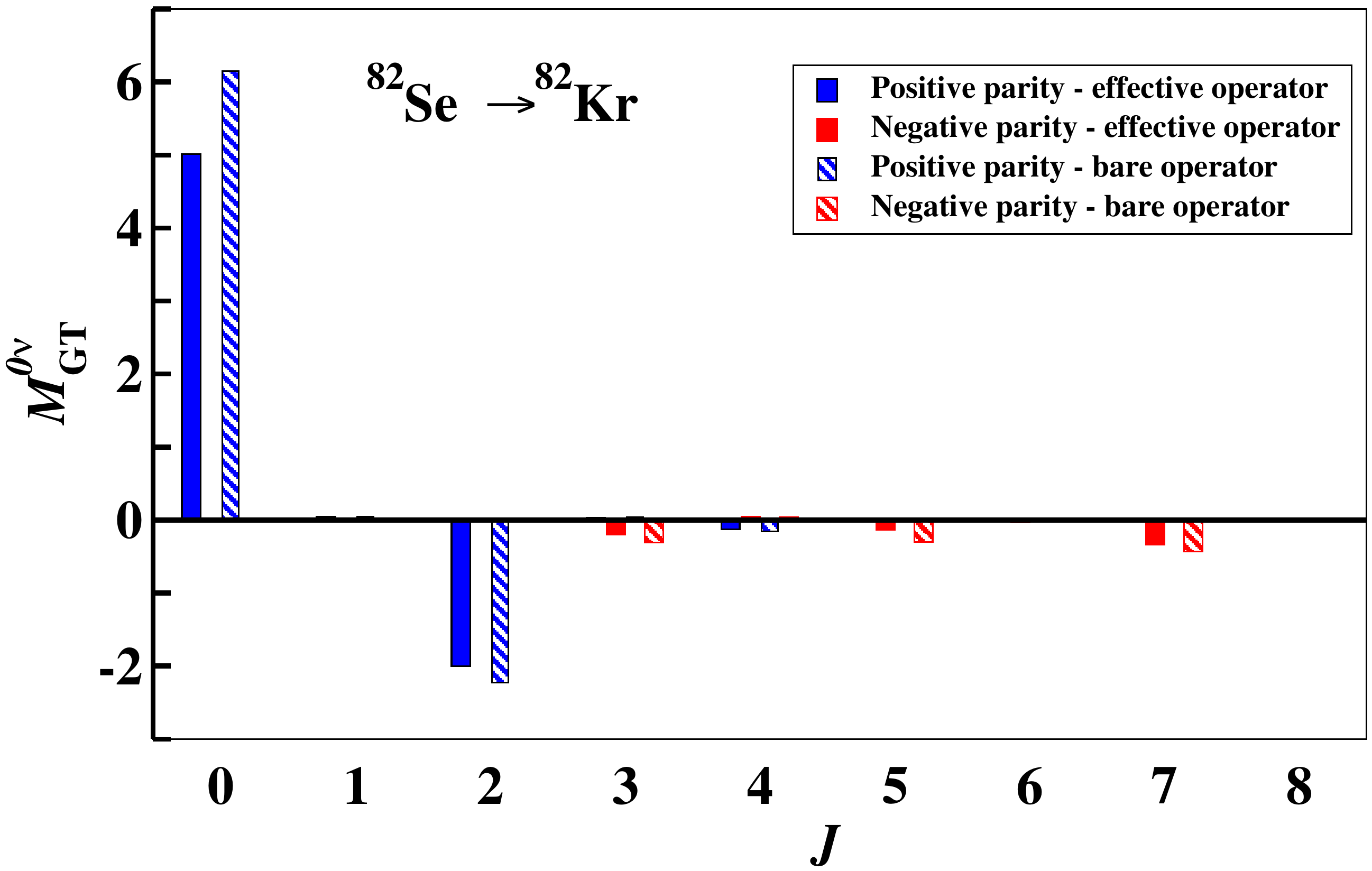}
\caption{Same as in Fig. \ref{76Ge-jj}, but for $^{82}$Se \zbb~decay.}
\label{82Se-jj}
\end{center}
\end{figure}

\begin{figure}[h]
\begin{center}
\includegraphics[scale=0.32,angle=0]{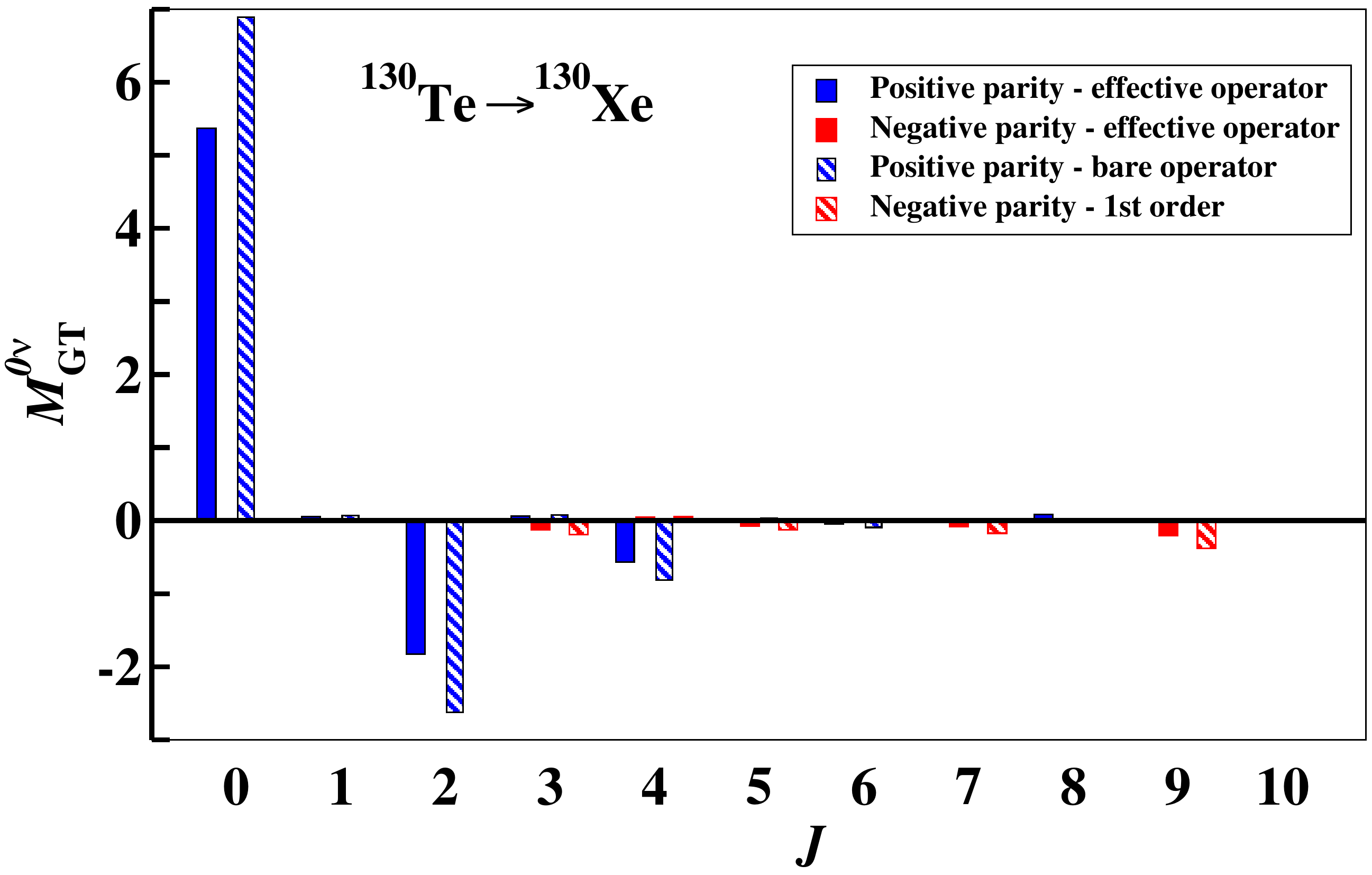}
\caption{Same as in Fig. \ref{76Ge-jj}, but for $^{130}$Te \zbb~decay.}
\label{130Te-jj}
\end{center}
\end{figure}

\begin{figure}[h]
\begin{center}
\includegraphics[scale=0.32,angle=0]{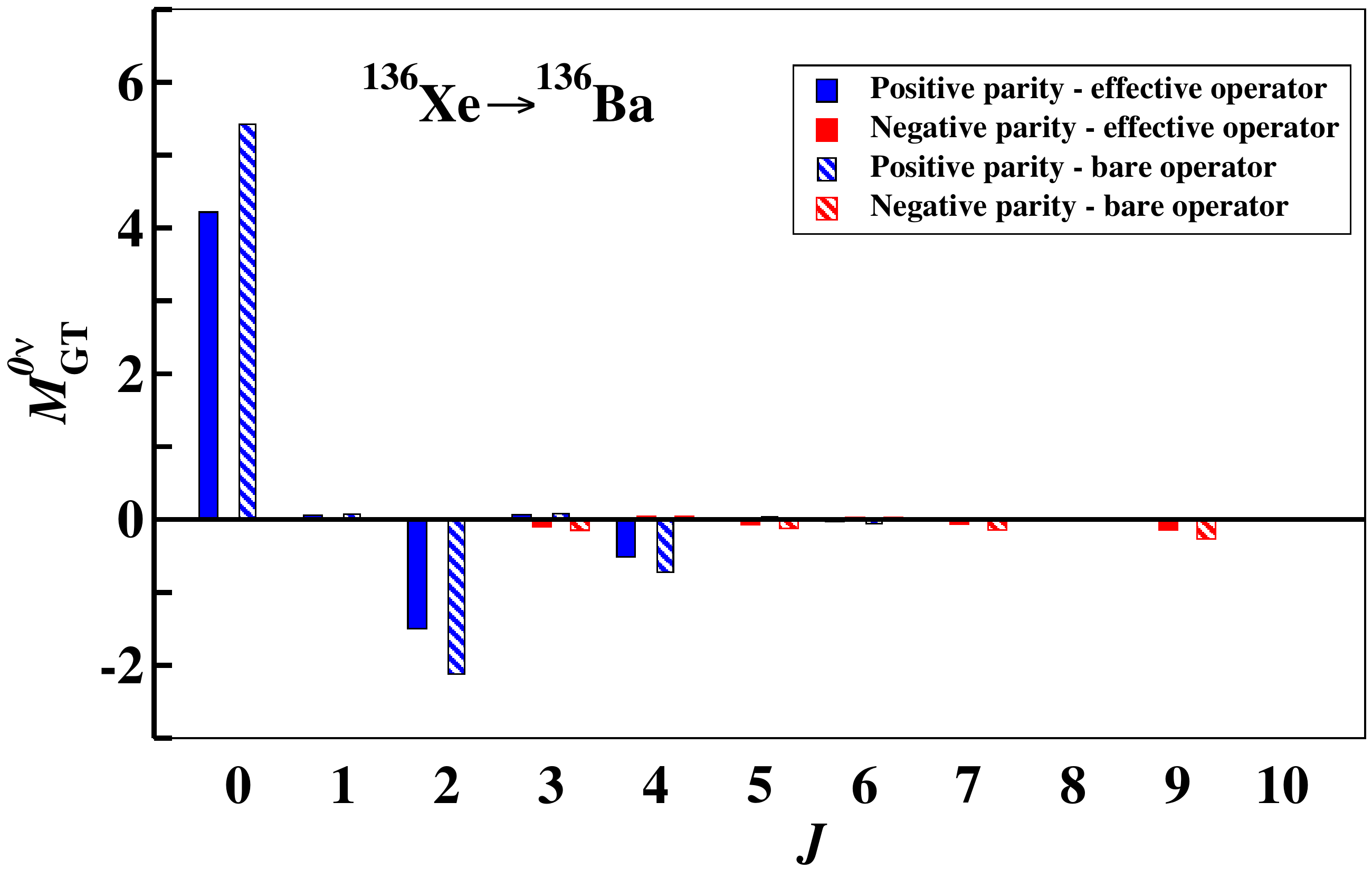}
\caption{Same as in Fig. \ref{76Ge-jj}, but for $^{136}$Xe \zbb~decay.}
\label{136Xe-jj}
\end{center}
\end{figure}

The calculation of \nme~for $^{48}$Ca \zbb~decay exhibits the worst
perturbative behavior.
In such a case, we observe a difference between the second- and
third-order results which is almost $50\%$, and this puzzling outcome
deserves a more specific discussion when the study of the GT matrix
elements in terms of the contributions from the decaying pair of
neutrons coupled to a given angular momentum and parity $J^{\pi}$ will
be reported in Figs. \ref{48Ca-jj}-\ref{136Xe-jj}.

The issue of the perturbative behavior needs to be addressed, and in
this connection the calculation of \nme~beyond the closure
approximation may lead to an improvement.
In fact, the energy denominator of neutrino potentials in
Eq. (\ref{neutpot}) depends on the energies of real intermediate
states, and the inclusion of this dependence in the calculation of
virtual intermediate states in the perturbative expansion of
$\Theta_{\rm eff}$ may strongly influence the order-by-order
perturbative behavior.

As mentioned before, we perform a decomposition of $M^{0\nu}_{\rm GT}$
in terms of the contributions from the decaying pair of neutrons
coupled to a given angular momentum and parity $J^{\pi}$, and the
results are reported in Figs. \ref{48Ca-jj}-\ref{136Xe-jj}.
We compare the contributions obtained from calculations performed by
employing both the effective \zbb-decay operator $\Theta_{\rm eff}$
and the bare one, without any renormalization contribution.

As a general remark, we see that each $J^{\pi}$ contribution to
$M^{0\nu}_{\rm GT}$ calculated employing $\Theta_{\rm eff}$ is smaller
than the one obtained with the bare \zbb-decay operator.
This corresponds to a quenching of each Gamow-Teller component, in
terms of the $J^{\pi}$ decomposition.
Actually, the main contributions, for both the effective and the bare
operators, are provided by $J^{\pi}=0^+,2^+$ components and they are
always opposite in sign.
A non-negligible role is also played by the $J^{\pi}=4^+$ component for
$^{130}$Te,$^{136}$Xe decays.

\begin{table}[ht]
  \caption{Calculated values of \nme~for all decays under
    investigation. The first column corresponds to the results
    ontained employing the bare \zbb-decay operator, the second one to
  the calculations performed with $\Theta_{\rm eff}$. The third column
  corresponds to calculating $\Theta_{\rm eff}$ without corrections
  due to the Pauli-blocking effect.}
\begin{ruledtabular}
\begin{tabular}{cccc}
\label{NME}
 Decay & bare operator & $\Theta_{\rm eff}$ & $\Theta_{\rm eff}$ -
                                                 no Pauli \\
\colrule
~ & ~ & ~& ~ \\
$^{48}$Ca  $\rightarrow$ $^{48}$Ti & 0.53 & 0.30 & 0.30 \\
$^{76}$Ge  $\rightarrow$ $^{76}$Se & 3.35 & 2.66 & 3.01 \\
$^{82}$Se  $\rightarrow$ $^{82}$Kr & 3.30 & 2.72 & 2.94 \\
$^{130}$Te $\rightarrow$ $^{130}$Xe & 3.27 & 3.16 & 2.98 \\
$^{136}$Xe  $\rightarrow$ $^{136}$Ba & 2.47 & 2.39 & 2.24 \\
\colrule
\end{tabular}
\end{ruledtabular}
\end{table}

The quenching of GT $J^{\pi}$ components reflects also on the
comparison between the full \nme~calculated with the bare and
effective \zbb-decay operator, as can be seen in the values reported
in Table \ref{NME}.
In the same table we also report the values of \nme~obtained without the
inclusion of the three-body contributions of Fig. \ref{figeffop3b},
which are intended to account for the Pauli exclusion principle in
many-valence-nucleons systems and correct the ``blocking effect''.

For the sake of the completeness, we point out that our results of the
$J^{\pi}$ decomposition are similar to those obtained in other
SM calculations, as for example in
Ref. \cite{Senkov13,Senkov16,Senkov14,Jiao18} for $^{48}$Ca,
$^{76}$Ge,  $^{82}$Se,  $^{130}$Te,  and $^{136}$Xe decays, respectively.

A long standing issue about the calculation of \nme~is
  wether there is a connection between the derivation of the effective
  one-body GT operator \cite{Coraggio19a} and the renormalization of
  the two-body GT component of the \zbb~operator, or not.
  Namely, some authors argue that the quenching factor that is needed
  to match theory and experiment of GT-decay properties 
  (single-$\beta$ decay strengths, \nmeds, etc.) should be also
  employed to calculate \nme, with a large impact on the detectability
  of \zbb~process (see for instance Refs. \cite{Suhonen17a,Suhonen17b}).

It is worth to observe that, from our results of the calculation of
\nmeds~using both bare and effective single-$\beta$ decay operators
(see Tables II, IV, VI, and VIII in Ref. \cite{Coraggio19a}), we may
induce quenching factors of the axial coupling constant $g_A$ that are
$q=0.83,0.58,0.56,0.68,0.61$ for $^{48}$Ca, $^{76}$Ge,
$^{82}$Se, $^{130}$Te, and $^{136}$Xe decays, respectively.

If these quenching factors were employed to calculate \nme, starting
from the bare \zbb-decay operator, we would obtain
\nme=0.40,1.41,1.32,1.78,1.15 for the above corresponding decays.

The comparison of these numbers with those in Table \ref{NME}
evidences the different mechanisms which underlies the renormalization
of the one-body single-$\beta$ and the two-body \zbb~decay operators,
the first one leading to quenching factors which might strongly
suppress the calculated \nmes.

Now, as mentioned above, it is worth to come back to the discussion
about the perturbative behavior of the effective \zbb -decay operator
for $^{48}\mbox{Ca}\rightarrow^{48}$Ti.
From the inspection of Fig. \ref{48Ca-jj}, we observe a large
cancelation between the $J^{\pi}=0^+$ and $J^{\pi}=2^+$ components,
leading to a \nme value that is the smallest among the nuclear matrix
elements we have calculated.
This is a characteristic that our results share with other SM
calculations such as those, for example, in
Refs. \cite{Menendez09b,Senkov13}.

\begin{figure}[h]
\begin{center}
\includegraphics[scale=0.32,angle=0]{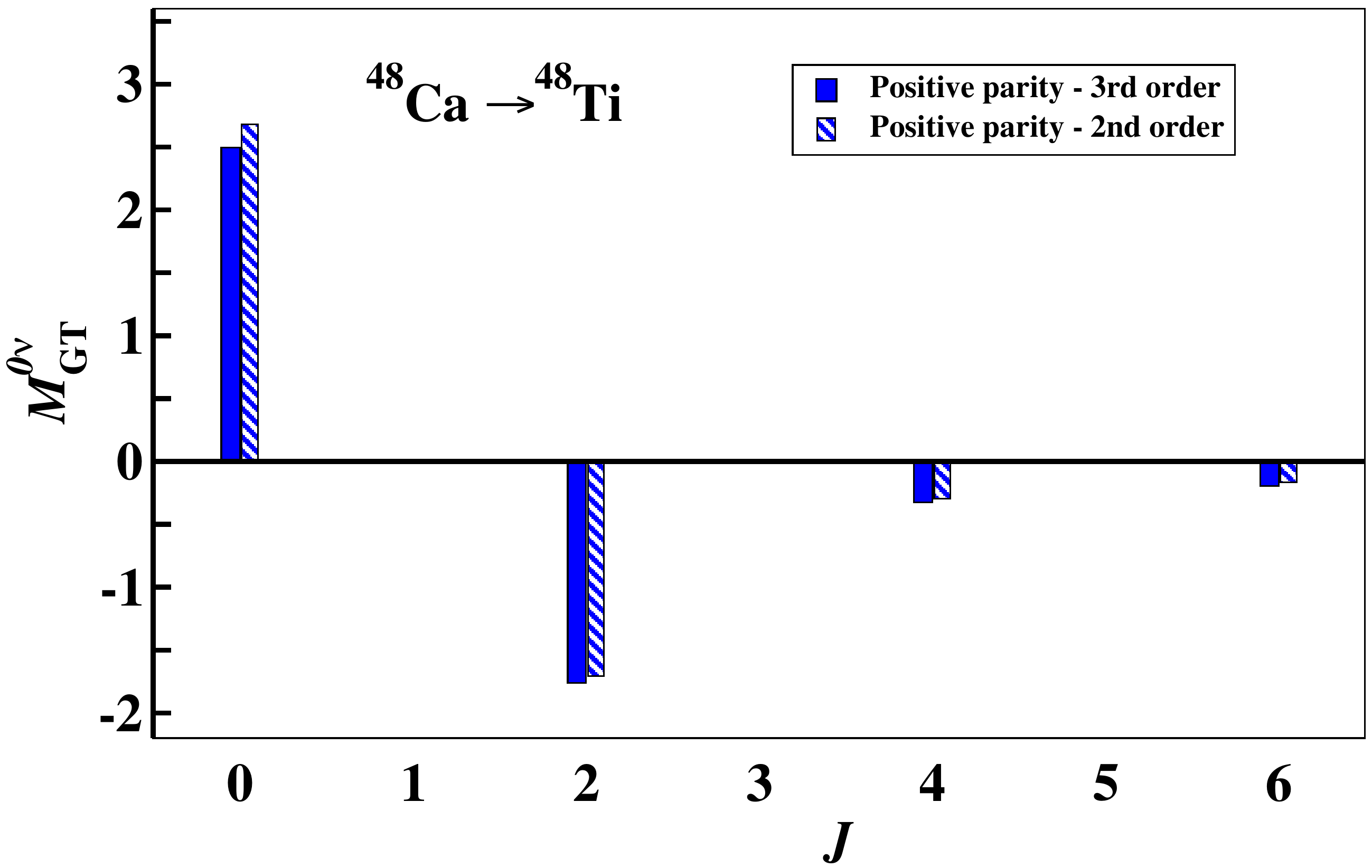}
\caption{Contributions from pairs of decaying neutrons with given
  $J^{\pi}$ to $M^{0\nu}_{\rm GT}$ for $^{48}$Ca \zbb~decay. The bars
  filled in blue corresponds to the results obtained with $\Theta_{\rm
    eff}$ calculated a third order in perturbation theory (as in
  Fig. \ref{48Ca-jj}), those in dashed blue are calculated at second
  order.}
\label{48Ca-jj-23}
\end{center}
\end{figure}

In Fig. \ref{48Ca-jj-23} we compare the results of the $J^{\pi}$
decomposition we obtain employing, for $^{48}$Ca decay, the effective
\zbb~operator calculated both at second and third order in
perturbation theory.

As can be seen, the order-by-order convergence of each component of
$M^{0\nu}_{\rm GT}$ is quite good, and the difference between second
and third order does not exceed $7\%$ for the two main contributions
$J^{\pi}=2^+,4^+$.
However, the cancelation between these two $J^{\pi}$ component
enhances the oscillation between the total result at second and third order in
perturbation theory, the difference between them increasing to $50\%$.

Finally, we would like to comment the comparison between the results
reported in the last two columns in Table \ref{NME}, that evidences the
role played by the ``blocking effect''.
As can be seen, the difference between the last two columns - results
with and without accounting for the Pauli-principle violations - is at
most about $12 \%$ for $^{76}$Ge decay, being less than $10\%$ in all
other decays.
In Ref. \cite{Coraggio19a} we have found that the contribution of the
``blocking effect'' in the \dbb~ decay has more or less the same
magnitude, but it always enlarges \nmed.
In the present calculations, we observe a decrease of \nme~for $^{76}$Ge
and $^{82}$Se \zbb~decays, and an increase for $^{130}$Te
and $^{136}$Xe \zbb~decays.
This feature is related to a different balance between the role of
diagrams (a) and (b) (see Fig. \ref{figeffop3b}) in the perturbative
expansion of $\Theta_{\rm eff}$, these diagrams suppressing the
Pauli-principle violating contribution of diagrams (a') and (b') (see
Table 3 in Ref. \cite{Itaco19}).

\section{Summary and outlook}\label{conclusions}
The work presented in this paper has been devoted to the calculation,
within the realistic shell model, of the nuclear matrix element
\nme~relative to the \zbb decay.
To this end, shell-model effective Hamiltonians and operators have
been derived by way of many-body perturbation theory, starting from a
realistic $NN$ potential renormalized through the \vlwk~approach
\cite{Bogner02}.

Our study has been focussed on the $^{48}$Ca$\rightarrow^{48}$Ti,
$^{76}$Ge$\rightarrow^{76}$Se, $^{82}$Se$\rightarrow^{82}$Kr,
$^{130}$Te$\rightarrow^{130}$Xe, and $^{136}$Xe$\rightarrow^{136}$Ba
decays, as a continuation of a previous work where we have addressed
the reliability of the realistic shell model to calculate, by way of
theoretical SM effective operators, the nuclear matrix elements
\nmeds~for the \dbb~ decay involving the same nuclei of present
investigation \cite{Coraggio19a}.
The results that have been obtained and shown in that paper support
the ability of realistic shell model to provide a quantitative
description of data relative to both spectroscopy (low-lying
excitation spectra, electromagnetic transition strengths) and $\beta$
decay (nuclear matrix elements of \dbb~decay, GT strengths from
charge-exchange reactions) without resorting to empirical adjustments
of \heff, effective charges or gyromagnetic factors, or to the
quenching of the axial coupling constant.

We may summarise the main results of present study into two
statements:
\begin{itemize}
\item the order-by-order perturbative behavior of the calculated
  effective two-body \zbb-decay operator is not as satisfactory as the
  one relative to the perturbative expansion of the effective one-body
  single-$\beta$ decay operator \cite{Coraggio18b}.
  The issue of the perturbative behavior of effective SM Hamiltonians
  and operators is relevant to assess the evaluation of theoretical
  uncertainties.
\item Even if our results may be still significantly improved, we can
  assert that the effect of the renormalization of the \zbb-decay
  operator, with respect to the truncation of the full Hilbert space to
  the shell-model one, is far less important than that regarding
  the well-known problem of the $g_A$ quenching
  \cite{Suhonen17b,Suhonen17a}.
  Namely, the \nmes~calculated using the effective
  \zbb-decay operator are quite larger than those calculated employing
  a quenching factor deduced by our results for the \dbb-decay (see
  Section \ref{calculations}).
\end{itemize}

Our present and near future efforts, on one side, are and will be
devoted to improve the perturbative behavior of the order-by-order
expansion of the effective \zbb~ operator.
As we have mentioned in the previous Section, we are confident that
overcoming the limits of the closure approximation this goal may be
achieved, even if this is a very demanding task in terms of
computational resources.

On another side, we intend to deal with the renormalization of
  the \zbb~ operator due to the subnucleonic degrees of freedom.
  This issue has to be consistently tackled starting from two- and
three-body nuclear potentials derived within the framework of chiral
perturbation theory \cite{Coraggio18a,Ma19}, and taking also into account
the contributions of chiral two-body electroweak currents.
Some authors have found that $\beta$- and
neutrinoless double-beta decays may be affected by these
contributions \cite{Gysbers19,Wang18}, even if it should be pointed
out that a recent study evidences that, for standard GT transitions
($\beta$ decay, \dbb~decay), the many-body meson
currents should play no significant role due to the ``chiral
filter'' mechanism \cite{Rho19a}.

Actually, this mechanism may be no longer valid for \zbb~decay, the
transferred momentum being $\sim 100$ MeV, and therefore
  we consider intriguing to carry out further investigations to grasp
  the role of exchange-current contributions.

At present, the derivation of effective SM Hamiltonians and operators
for medium-mass nuclei - including also contributions of three-body
nuclear potentials and many-body electroweak currents - is
computationally challenging if one wants to achieve the same accuracy
attained in present calculations, namely carrying out the perturbative
expansion consistently up to the third order in perturbation theory.

\vspace{0.5truecm}

\section*{Acknowledgements}
The authors thank Prof. Mannque Rho for a valuable and fruitful discussion about
the role of many-body electroweak currents.
\bibliographystyle{apsrev}
\bibliography{biblio.bib}

\end{document}